\documentclass[preprint]{aastex63}
\usepackage{amsmath}
\usepackage{mathtools}
\usepackage{bm}

\received{--}
\revised{--}
\accepted{--}
\submitjournal{ApJS}

\shorttitle{Extension of Athena++ for Conservative Self-Gravity}
\shortauthors{Mullen et al.}
\graphicspath{{./}{figures/}}
\begin{document}

\title{An Extension of the Athena++ Framework for Fully Conservative Self-Gravitating Hydrodynamics}

\correspondingauthor{P. D. Mullen}
\email{pmullen2@illinois.edu}

\author[0000-0003-2131-4634]{P. D. Mullen}
\affiliation{Department of Astronomy, University of Illinois at Urbana-Champaign, 1002 West Green Street, Urbana, IL, 61801, USA}

\author[0000-0002-7538-581X]{Tomoyuki Hanawa}
\affiliation{Center for Frontier Science, Chiba University, 1-33 Yayoi-cho, Inage-ku, Chiba 263-8522, Japan}

\author[0000-0001-7451-8935]{C. F. Gammie}
\affiliation{Department of Astronomy, University of Illinois at Urbana-Champaign, 1002 West Green Street, Urbana, IL, 61801, USA}
\affiliation{Department of Physics, University of Illinois at Urbana-Champaign, 1110 West Green Street, Urbana, IL 61801, USA} 



\begin{abstract}
Numerical simulations of self-gravitating flows evolve a momentum equation and an energy equation that account for accelerations and gravitational energy releases due to a time-dependent gravitational potential.  In this work, we implement a fully conservative numerical algorithm for self-gravitating flows, using source terms, in the astrophysical magnetohydrodynamics framework \texttt{Athena++}.  We demonstrate that properly evaluated source terms are conservative when they are equivalent to the divergence of a corresponding ``gravity flux" (i.e., a gravitational stress tensor or a gravitational energy flux). We provide test problems that demonstrate several advantages of the source-term-based algorithm, including second order convergence and round-off error total momentum and total energy conservation.  The fully conservative scheme suppresses anomalous accelerations that arise when applying a common numerical discretization of the gravitational stress tensor that does not guarantee curl-free gravity.  
\end{abstract}



\section{Introduction} \label{sec:introduction}
Self-gravity is dynamically important in many astrophysical settings: core collapse supernovae explosions \citep[e.g.,][]{nordhaus+2010,couch+2013},  Moon-forming giant impacts \citep[e.g.,][]{thompson+1988,wada+2006,canup+2013}, planet formation \citep[e.g.,][]{boss1997,rice+2005,simon+2016}, star formation \citep[e.g.,][]{ostriker+2001,mckee+2007}, and white dwarf mergers \citep[e.g.,][]{katz+2016}, to name a few.  Self-gravitating astrophysical dynamics are often physically complex, with gravity interacting with a diversity of other physics (e.g., magnetic fields, radiation, non-ideal equations of state, etc.), hence, to better understand such complicated systems, numerical simulations are often employed.  

Numerical simulations of self-gravitating astrophysical systems must evolve a flow subject to a time-dependent gravitational potential specified by the Poisson equation.  The need for fast, accurate evaluations of the gravitational potential have inspired the development of elliptical solvers employing, for example, Fast-Fourier transforms  \citep[e.g.,][]{hockney+1988,moon+2019} or multigrid methods \citep[e.g.,][Tomida et al., in preparation]{matsumoto+2003,ricker2008}. 

Equally as important, special care must be given to the numerical evaluation of gravitational accelerations and gravitational energy releases acting on a flow.   Various algorithms have therefore been proposed to integrate the momentum and energy equations for self-gravitating hydrodynamics.  \cite{jiang+2013} presented a fully conservative numerical scheme (i.e., a ``gravity flux" scheme) that evolves the momentum equation and total energy equation by evaluating the divergence of the gravitational stress tensor and a gravitational ``energy flux", respectively.  They argue that this scheme excels in maintaining the shape/equilibria of self-gravitating systems, particularly upon advection \citep{jiang+2013}.  Developments in \cite{mikami+2008}, \cite{springel2010}, \cite{katz+2016}, and \cite{hanawa2019} instead argue in favor of a source-term based approach to evaluate the momentum and energy equations, as the ``gravity flux" scheme can produce significant errors in gravitational accelerations when the density implied by the discretized Poisson equation differs from the local density.  Moreover, we find that a common discretization of the gravitational stress tensor in the ``gravity flux" scheme \citep[e.g.,][]{stone+2008,stone+2020} can produce gravitational accelerations that are not curl-free, hence producing anomalous behavior in regions with low density and large gravity.  In particular, in a low mass medium (e.g., a disk, atmosphere, or ambient background) where the gravity is dominated by a body of mass $M$ and characteristic length scale $L$, the magnitude of the anomalous accelerations may be comparable to the true gravitational accelerations when the mass density $\rho < M/L^3 \left( \Delta x / L \right)^2$ where $\Delta x$ is the numerical linear resolution (see later \S{\ref{sec:curl_g_zero}}).

Seemingly, the choice of numerical scheme employed to integrate the momentum and energy equations may come with distinct advantages, but also, potential drawbacks.  In this work, we follow developments in \cite{mikami+2008}, \cite{springel2010}, \cite{katz+2016}, and \cite{hanawa2019} and implement a fully conservative scheme, using source terms, for self-gravitating hydrodynamics in the \texttt{Athena++} framework  \citep{stone+2020}.  In \S\ref{sec:fully_conservative_source_terms}, we present a proof demonstrating that source terms, when evaluated properly, can have a corresponding flux.  This equivalence guarantees total momentum and total energy conservation.  In \S\ref{sec:algorithm_properties_advantages}, we describe how the fully conservative source term scheme can be implemented into a numerical hydrodynamics framework; we show several of its key properties/advantages: second order accuracy in space and time, the requirement of only two Poisson solves per numerical time-step (for temporally second-order accurate time integrators, see however Appendix \S\ref{sec:rk3}), and total momentum and total energy conservation to round-off error.  We highlight the scheme's ability to suppress anomalous accelerations that arise when applying a common numerical discretization of the gravitational stress tensor in \S{\ref{sec:curl_g_zero}}.  In \S\ref{sec:implementation_in_athena++}, we rigorously test an implementation of the fully conservative source term algorithm in \texttt{Athena++} via multiple test problems, including Spitzer sheets (1-D equilibria), Jeans linear waves (in 3-D), and polytropic equilibria (with low-mass overlying atmospheres).  In \S\ref{sec:discussion} and \S\ref{sec:conclusion}, we provide a discussion and conclusion. 

\section{Fully Conservative Source Terms} \label{sec:fully_conservative_source_terms}
\subsection{Governing Equations} \label{sec:governing_equations}
The Eulerian equations of self-gravitating hydrodynamics evolve a flow's spatially varying density $\rho$, velocity $\mathbf{v}$, and pressure $P$ subject to a time-dependent gravitational potential $\phi$ obeying the Poisson equation, 
\begin{equation} \label{eq:poisson}
\nabla^2 \phi = 4 \pi G \rho,
\end{equation}
where $G$ is the gravitational constant.  The continuity, momentum, and energy equations can be expressed in non-conservative form as
\begin{equation}
\frac{\partial \rho}{\partial t} + \nabla \cdot \left[\rho \mathbf{v} \right]  =  0, \label{eq:continuity}
\end{equation}
\begin{equation}
\frac{\partial \rho \mathbf{v}}{\partial t} + \nabla \cdot \left[ \rho \mathbf{v} \mathbf{v} + \mathbf{P} \right]  =  \rho \mathbf{g}, \label{eq:momentum_eq_src}
\end{equation}
\begin{equation}
\frac{\partial E}{\partial t} + \nabla \cdot \left[ \left( E+P \right) \mathbf{v} \right]  =  \rho \mathbf{v} \cdot \mathbf{g}, \label{eq:energy_eq_src}
\end{equation}
where $\rho \mathbf{v}$ is the momentum density, $E$ is the energy density
\begin{equation}
E = e + \frac{\rho \left( \mathbf{v} \cdot \mathbf{v} \right)}{2},
\end{equation}
$e$ is the internal energy density, and $\mathbf{g}$ is the gravitational acceleration 
\begin{equation}
\mathbf{g} = -\nabla \phi,
\end{equation}
subject to the constraint
\begin{equation}
    \nabla \times \mathbf{g} = 0.
    \label{eq:curl}
\end{equation}

Equations (\ref{eq:momentum_eq_src}) and (\ref{eq:energy_eq_src}) are not unique.  Alternatively, \cite{jiang+2013} identified that the momentum and energy equations can instead be rewritten in fully conservative forms (i.e., where gravity source terms are recast as ``gravity fluxes")
\begin{equation}
\frac{\partial \rho \mathbf{v}}{\partial t} + \nabla \cdot \left[ \rho \mathbf{v} \mathbf{v} + \mathbf{P} + \mathbf{T_g} \right] = 0, \label{eq:momentum_eq_flx}
\end{equation}
\begin{equation}
\frac{\partial}{\partial t} \left(E + E_\mathrm{g} \right) + \nabla \cdot \left[(E+P)\mathbf{v} + \mathbf{F_g} \right] = 0, \label{eq:energy_eq_flx}
\end{equation}
where $\mathbf{T_g}$ is the gravitational stress tensor
\begin{equation}
    \mathbf{T_g} = \frac{1}{4 \pi G} \left[\nabla \phi \nabla \phi - \frac{1}{2} \left( \nabla \phi \right) \cdot \left( \nabla \phi \right) \mathbf{I} \right], 
\end{equation}
$\mathbf{I}$ is the identity tensor, $E_\mathrm{g}$ is the canonical gravitational energy density for a self-gravitating system,
\begin{equation} \label{eq:egform1}
E_\mathrm{g} = \frac{1}{2} \rho \phi,
\end{equation}
$\mathbf{F_g}$ is the gravitational ``energy flux" 
\begin{equation}
\mathbf{F_g} = \frac{1}{8 \pi G} \left(\phi \nabla \dot{\phi} - \dot{\phi} \nabla \phi \right) + \rho \mathbf{v} \phi,
\end{equation}

\noindent and $\dot{\phi} = \partial \phi/ \partial t$. 

\cite{hanawa2019} recognized that there exists yet another conservative form of the energy equation
\begin{equation} \label{eq:alt_energy_flux}
\frac{\partial}{\partial t} \left(E - \frac{\mathbf{g} \cdot \mathbf{g}}{8 \pi G} \right) 
+ \mathbf{\nabla} \cdot \left[ \left( E + P \right) \mathbf{v} + \mathbf{F_g^\prime} \right] = 0,
\end{equation}
where
\begin{equation}
\mathbf{F_g^\prime} = - \frac{\phi}{4\pi G}
\frac{\partial \mathbf{g}}{\partial t} + \rho \mathbf{v} \phi
\end{equation}
represents an alternative form of the gravitational energy flux. 
Equation (\ref{eq:alt_energy_flux}) is equivalent to Equation (\ref{eq:energy_eq_flx}) since
\begin{eqnarray}
- \frac{\partial}{\partial t} \left( \frac{\mathbf{g} \cdot \mathbf{g}}{8 \pi G} \right) & = &
- \frac{1}{4 \pi G} \left[\mathbf{g} \cdot \frac{\partial \mathbf{g}}{\partial t} \right] \label{eq:proper2avg} \\ 
& = & \mathbf{\nabla} \cdot \left[ \frac{\phi}{4 \pi G} \frac{\partial \mathbf{g}}{\partial t} \right]
+ \phi \frac{\partial \rho}{\partial t} \label{eq:proper2sub} \\
& = & \mathbf{\nabla} \cdot \left[ \frac{\mathbf{g}}{4 \pi G} \frac{\partial \phi}{\partial t} \right]
+ \rho \frac{\partial \phi}{\partial t} \\
& = & \frac{\partial}{\partial t} \left[ \frac{\rho \phi}{2}
+ \mathbf{\nabla} \cdot \left( \frac{\phi \mathbf{g}}{8\pi G} \right) \right].
\label{eq:dteg}
\end{eqnarray}
Upon inspection, Equation (\ref{eq:alt_energy_flux}) gives an alternative form of the gravitational energy density for self-gravitating systems:
\begin{equation}
E _{\mathrm{g}}^\prime = -\frac{\mathbf{g} \cdot \mathbf{g}}{8 \pi G} .   \label{eq:egform2}
\end{equation}
The volume-integrated ($\int dV$) gravitational energy densities from Equation (\ref{eq:egform1}) and (\ref{eq:egform2}) are equivalent
\begin{equation}
\begin{aligned}
\int E_\mathrm{g} dV = \int \frac{\rho \phi}{2} dV & = {} \hphantom{-} \int \frac{ \phi \nabla^2 \phi}{8 \pi G} dV  \\
& =  - \int \frac{\mathbf{g} \cdot \mathbf{g}}{8 \pi G} dV -  \int \left( \frac{ \phi \mathbf{g}}{8 \pi G} \right)  \cdot
d\mathbf{S} , \\ 
& = \hphantom{-} \int E_\mathrm{g}^\prime dV -  \int \left( \frac{ \phi \mathbf{g}}{8 \pi G} \right)  \cdot
d\mathbf{S}
\end{aligned}
\label{eq:volume_int_equal}
\end{equation}
when the surface integral $\int d\mathbf{S}$ vanishes, i.e., when periodic boundary conditions are applied or when
most of mass is concentrated in regions far from the outer boundary.

\subsection{Equivalence} \label{sec:equivalence}
The non-conservative (Equations \ref{eq:momentum_eq_src} and \ref{eq:energy_eq_src}) and conservative (Equations \ref{eq:momentum_eq_flx}, \ref{eq:energy_eq_flx}, and \ref{eq:alt_energy_flux}) formulations of the momentum and energy equations lend themselves to two entirely differently numerical algorithms. A numerical implementation integrating Equations (\ref{eq:momentum_eq_src}-\ref{eq:energy_eq_src}) applies time-explicit source terms to the momentum density $\rho \mathbf{v}$ and energy density $E$.  In contrast, a numerical scheme evaluating Equations (\ref{eq:momentum_eq_flx}, \ref{eq:energy_eq_flx}, and \ref{eq:alt_energy_flux}) requires computing gravitational momentum fluxes (i.e., the gravitational stress tensor $\mathbf{T_g}$) and gravitational energy fluxes (i.e., $\mathbf{F_g}$ or $\mathbf{F_g^\prime}$).  By taking the numerical divergence of fluxes, the latter ``gravity flux" scheme guarantees conservation of total momentum and total energy
\begin{equation}
\frac{\partial}{\partial t} \int \left(\rho \mathbf{v} \right) dV =  0,
\end{equation}
\begin{equation}
\frac{\partial}{\partial t} \int \left(E+ E_\mathrm{g} \right) dV = \frac{\partial}{\partial t} \int \left(E+ E_\mathrm{g}^\prime \right) dV = 0,
\end{equation}
to numerical round-off error when periodic boundary conditions are applied.

Typically, obtaining round-off error total momentum and total energy conservation is not possible when using a source-term based approach, however, following developments in \cite{mikami+2008}, \cite{springel2010}, \cite{katz+2016}, and \cite{hanawa2019}, we now demonstrate that if the source terms in Equations (\ref{eq:momentum_eq_src}) and (\ref{eq:energy_eq_src}) are constructed such that they are equivalent to corresponding gravitational fluxes, then the source term approach can be fully conservative.  By equating Equation (\ref{eq:momentum_eq_src}) with Equation (\ref{eq:momentum_eq_flx}) and Equation (\ref{eq:energy_eq_src}) with Equation (\ref{eq:alt_energy_flux}), we identify the necessary equivalences
\begin{eqnarray}
\rho \mathbf{g} & = & -\nabla \cdot \mathbf{T_g} , \label{eq:proper1} \\
\rho \mathbf{v} \cdot \mathbf{g} & = & - \mathbf{\nabla} \cdot \mathbf{F_g^\prime} + \frac{\partial}{\partial t} \left(\frac{\mathbf{\mathbf{g} \cdot \mathbf{g}}}{8 \pi G} \right) \label{eq:proper2}.
\end{eqnarray}

\subsection{Properly Evaluated Source Terms} \label{sec:proper_src}
In the following, we seek finite difference equations which satisfy Equations (\ref{eq:proper1}) and (\ref{eq:proper2}) for each numerical cell. Consider a uniform rectangular grid in Cartesian coordinates where the position of the cell center is designated
\begin{equation}
\left( x_i, y _j, z _k \right) = \left( i \Delta x, j \Delta y, k \Delta z \right) .
\end{equation}
Here, the indices, $ i $, $ j $, and $ k $ denote the cell number in the $ x $-, $ y $- and
$ z $-directions, respectively.   The grid spacings, $ \Delta x $, $ \Delta y $, and 
$ \Delta z $, can be either equal or different.  Using centered differences, we 
discretize the Poisson equation (Equation \ref{eq:poisson}) as
\begin{eqnarray}
\begin{aligned}
\frac{\phi _{i+1,j,k} - 2 \phi _{i,j,k} + \phi _{i-1,j,k}}{\Delta x^2} & + \frac{\phi _{i,j+1,k} - 2 \phi _{i,j,k} + \phi _{i,j-1,k}}{\Delta y^2} \\
& + \frac{\phi _{i,j,k+1} - 2 \phi _{i,j,k} + \phi _{i,j,k-1}}{\Delta z^2} = 4 \pi G \rho _{i,j,k}
\end{aligned}
\label{eq:poisson_discretized}
\end{eqnarray}
where $ \phi _{i,j,k} $ and $ \rho _{i,j,k} $ denote the gravitational potential and
the density at the cell center, respectively.  Equations (\ref{eq:momentum_eq_flx}) and (\ref{eq:alt_energy_flux}) demonstrate that gravitational accelerations and gravitational energy releases arise from taking the divergence of the gravitational stress tensor and an ``energy flux", hence, the source terms should have the gravity $\mathbf{g}$ defined at cell faces,
\begin{eqnarray}
\mathrm{g} _{x,i+1/2,j,k} & = & - \frac{\phi _{i+1,j,k} - \phi _{i,j,k}}{\Delta x} ,  \label{eq:gx}\\
\mathrm{g} _{y,i,j+1/2,k} & = & - \frac{\phi _{i,j+1,k} - \phi _{i,j,k}}{\Delta y} ,  \label{eq:gy} \\
\mathrm{g} _{z,i,j,k+1/2} & = &- \frac{\phi _{i,j,k+1} - \phi _{i,j,k}}{\Delta z}. \label{eq:gz}
\end{eqnarray}
By use of Equations (\ref{eq:gx}-\ref{eq:gz}), Equation (\ref{eq:poisson_discretized}) is rewritten as
\begin{eqnarray}
\begin{aligned}
- \frac{\mathrm{g} _{x,i+1/2,j,k} - \mathrm{g} _{x,i-1/2,j,k}}{\Delta x} & 
- \frac{\mathrm{g} _{y,i,j+1/2,k} - \mathrm{g} _{y,i,j-1/2,k}}{\Delta y} \\ 
& - \frac{\mathrm{g} _{z,i,j,k+1/2} - \mathrm{g} _{z,i,j,k-1/2}}{\Delta z} = 4 \pi G \rho _{i,j,k}  .
\end{aligned}
\label{eq:poisson_discretized2}
\end{eqnarray}
Multiplying $ \phi _{i,j,k} \Delta x \Delta y \Delta z / (8 \pi G) $ to Equation (\ref{eq:poisson_discretized2})
we obtain
\begin{eqnarray}
E_\mathrm{g} \Delta x \Delta y \Delta z \; & = & \hphantom{-}E_\mathrm{g}^\prime \Delta x \Delta y \Delta z \nonumber \\
& & - \frac{\mathrm{g} _{x,i+1/2,j,k} \left( \phi _{i+1,j,k} + \phi _{i,j,k} \right)}{16 \pi G} \Delta y \Delta z
+ \frac{\mathrm{g} _{x,i-1/2,j,k} \left( \phi _{i,j,k} + \phi _{i-1,j,k} \right)}{16 \pi G} \Delta y \Delta z  \nonumber \\
& & - \frac{\mathrm{g} _{y,i,j+1/2,k} \left( \phi _{i,j+1,k} + \phi _{i,j,k} \right)}{16 \pi G} \Delta z \Delta x
+ \frac{\mathrm{g} _{y,i,j-1/2,k} \left( \phi _{i,j,k} + \phi _{i,j-1,k} \right)}{16 \pi G} \Delta z \Delta x  \nonumber \\
& & - \frac{\mathrm{g} _{z,i,j,k+1/2} \left( \phi _{i,j,k+1} + \phi _{i,j,k} \right)}{16 \pi G} \Delta x \Delta y
+ \frac{\mathrm{g} _{z,i,j,k-1/2} \left( \phi _{i,j,k} + \phi _{i,j,k-1} \right)}{16 \pi G} \Delta x \Delta y.
\label{eq:poisson_discretized3}
\end{eqnarray}
Equation (\ref{eq:poisson_discretized3}) means that the equality $\int E _\mathrm{g} \; dV = \int E _\mathrm{g} ^\prime \; dV $ holds if the gravitational energies are defined as
\begin{equation}
\int E _\mathrm{g} \; dV =  \frac{1}{2} \sum _{i,j,k} \phi _{i,j,k} \rho _{i,j,k} \Delta x \Delta y \Delta z ,
\label{eq:eg_sum_discretized}
\end{equation}
\begin{equation}
\int E _\mathrm{g}^\prime \; dV =  -\frac{1}{8 \pi G} \sum_{i,j,k} \frac{1}{2}
\left[
\begin{aligned}
{} 
&  \hphantom{\: +} \left( \mathrm{g} _{x,i-1/2,j,k} \right)^2 
+\left( \mathrm{g} _{x,i+1/2,j,k} \right)^2 \\ & +\left( \mathrm{g} _{y,i,j-1/2,k} \right)^2 
+\left( \mathrm{g} _{y,i,j+1/2,k} \right)^2 \\ &
+\left( \mathrm{g} _{z,i,j,k-1/2} \right)^2
+\left( \mathrm{g} _{z,i,j,k+1/2} \right)^2
\end{aligned} \right] \Delta x \Delta y \Delta z, \label{eq:egp_sum_discretized}
\end{equation}
and the surface terms are negligibly small.

\subsubsection{The Momentum Source Term} \label{sec:momentum_source_term}
The divergence of the gravitational stress tensor $\mathbf{T_g}$ gives
\begin{eqnarray}
-\mathbf{\nabla \cdot T} _{\rm g} & = & -\frac{\left( \mathbf{\nabla} \cdot \mathbf{g}\right)}{4\pi G} \mathbf{g} -
\frac{\left( \mathbf{\nabla} \times \mathbf{g} \right)}{4\pi G} \times \mathbf{g} \label{eq:tg_proof},
\end{eqnarray}
where the final term should vanish due to the curl-free constraint on the gravity $\mathbf{g}$ in (Equation \ref{eq:curl}).  Note that only the gravity normal to the cell surface appears in the discretized Poisson equation (Equation \ref{eq:poisson_discretized2}).  By extension, only normal components of the gravity should be used when computing the components of the gravitational stress tensor.  In prior work \citep[e.g.,][]{stone+2008,stone+2020}, the discretized gravitational stress tensor $\mathbf{\tilde{T}_g}$ has been computed as (for brevity only three components are shown)
\begin{eqnarray}
\tilde{T}_{xx,i+1/2,j,k} & = & \frac{(\mathrm{g}_{x,i+1/2,j,k})^2}{8 \pi G} - \frac{(\mathrm{g}_{y,i+1,j+1/2,k}+\mathrm{g}_{y,i,j+1/2,k}+\mathrm{g}_{y,i+1,j-1/2,k}+\mathrm{g}_{y,i,j-1/2,k})^2}{128 \pi G} \label{eq:Tg_tildexx} \\ 
& & - \frac{(\mathrm{g}_{z,i+1,j,k+1/2} + \mathrm{g}_{z,i,j,k+1/2} + \mathrm{g}_{z,i+1,j,k-1/2} + \mathrm{g}_{z,i,j,k-1/2})^2}{128 \pi G} \\
\tilde{T}_{yx,i,j+1/2,k} & = & \frac{\mathrm{g} _{y,i,j+1/2,k} ( \mathrm{g} _{x,i+1/2,j+1,k} + \mathrm{g}  _{x,i+1/2,j,k}
+ \mathrm{g} _{x,i-1/2,j+1,k} + \mathrm{g} _{x,i-1/2,j,k})}{16\pi G} \label{eq:Tg_tildeyx} \\
\tilde{T}_{zx,i,j,k+1/2} & = & \frac{\mathrm{g} _{z,i,j,k+1/2} ( \mathrm{g} _{x,i+1/2,j,k+1} + \mathrm{g}  _{x,i+1/2,j,k}
+ \mathrm{g} _{x,i-1/2,j,k+1} + \mathrm{g} _{x,i-1/2,j,k})}{16\pi G} \label{eq:Tg_tildezx}, 
\end{eqnarray}
As we shall later see (\S\ref{sec:curl_g_zero} and \S\ref{sec:polytropes}), this prescription for the discretized gravitational stress tensor, albeit second-order accurate, does not guarantee that the gravity obeys the $\nabla \times \mathbf{g} = 0$ constraint and can yield significant anomalous accelerations for problems with large density/mass contrasts.  Therefore, in this work, we advocate for another discretization of the gravitational stress tensor $\mathbf{T_g}$ where components are defined following
\begin{eqnarray}
T _{xx,i+1/2,j,k} & = & \frac{\left( \mathrm{g} _{x,i+1/2,j,k} \right)^2}{8\pi G}
- \frac{\mathrm{g} _{y,i+1,j+1/2,k} \mathrm{g} _{y,i,j+1/2,k} + \mathrm{g} _{y,i+1,j-1/2,k} \mathrm{g} _{y,i,j-1/2,k}}{16 \pi G} 
\label{eq:Tg_xx} \\
& & - \frac{\mathrm{g} _{z,i+1,j,k+1/2} \mathrm{g} _{z,i,j,k+1/2} + \mathrm{g} _{z,i+1,j,k-1/2} \mathrm{g} _{z,i,j,k-1/2}}{16 \pi G} , \\
T _{yx,i,j+1/2,k} & = & \tilde{T} _{yx,i,j+1/2,k}, \\
T _{zx,i,j,k+1/2} & = & 
\tilde{T} _{zx,i,j,k+1/2} ,
\end{eqnarray}

Using this new discretization of the gravitational stress tensor $\mathbf{T_g}$, we show in Appendix \ref{sec:divergence_of_the_grav_stress_tensor} that the fully conservative, discretized momentum source terms in Equation (\ref{eq:proper1}) simplify to
\begin{eqnarray}
\left( \rho \mathbf{g} \right) _{x,i,j,k} & = & 
\rho _{i,j,k} \; \mathrm{g}_{x,i,j,k} ,  \label{eq:proper_momentum_discretized_x} \\
\left( \rho \mathbf{g} \right) _{y,i,j,k} & = & 
\rho _{i,j,k} \; \mathrm{g}_{y,i,j,k},  \label{eq:proper_momentum_discretized_y} \\
\left( \rho \mathbf{g} \right) _{z,i,j,k} & = & 
\rho _{i,j,k} \; \mathrm{g}_{z,i,j,k},  \label{eq:proper_momentum_discretized_z}
\end{eqnarray}
\noindent where
\begin{eqnarray}
\mathrm{g}_{x,i,j,k} & = & \left(\mathrm{g}_{x,i+1/2,j,k} + \mathrm{g}_{x,i-1/2,j,k} \right)/2, \\ 
\mathrm{g}_{y,i,j,k} & = & \left(\mathrm{g}_{y,i,j+1/2,k} + \mathrm{g}_{y,i,j-1/2,k} \right)/2, \\
\mathrm{g}_{z,i,j,k} & = & \left(\mathrm{g}_{z,i,j,k+1/2} + \mathrm{g}_{z,i,j,k-1/2} \right)/2
\end{eqnarray}
are the three components of the cell-centered gravity, computed from the average of interface normal components.  The gravity $\mathbf{g}$ appearing in the properly evaluated source terms (Equations \ref{eq:proper_momentum_discretized_x}-\ref{eq:proper_momentum_discretized_z}) upholds the $\nabla \times \mathbf{g} = 0$ constraint, where
\begin{eqnarray}
(\nabla \times \mathbf{g})_{x,i,j,k} & = & \frac{\mathrm{g}_{z,i,j+1,k} - \mathrm{g}_{z,i,j-1,k}}{2 \Delta y} - \frac{\mathrm{g}_{y,i,j,k+1} - \mathrm{g}_{y,i,j,k-1}}{2 \Delta z}, \label{eq:curlgx} \\
(\nabla \times \mathbf{g})_{y,i,j,k} & = & \frac{\mathrm{g}_{x,i,j,k+1} - \mathrm{g}_{x,i,j,k-1}}{2 \Delta z} - \frac{\mathrm{g}_{z,i+1,j,k} - \mathrm{g}_{z,i-1,j,k}}{2 \Delta x}, \label{eq:curlgy} \\
(\nabla \times \mathbf{g})_{z,i,j,k} & = & \frac{\mathrm{g}_{y,i+1,j,k} - \mathrm{g}_{y,i-1,j,k}}{2 \Delta x} - \frac{\mathrm{g}_{x,i,j+1,k} - \mathrm{g}_{x,i,j-1,k}}{2 \Delta y}. \label{eq:curlgz}
\end{eqnarray}
\subsubsection{The Energy Source Term} \label{sec:energy_source_term}
After substituting Equation (\ref{eq:proper2sub}), we can recast the equivalence in Equation (\ref{eq:proper2}) as
\begin{equation}
\rho \mathbf{v} \cdot \mathbf{g} = - \nabla \cdot \left[\rho \mathbf{v} \phi \right] - \phi \frac{\partial \rho}{\partial t} \label{eq:proper2alt}
\end{equation}
The right hand side of Equation (\ref{eq:proper2alt}) is evaluated to be
\begin{eqnarray}
\left( \rho \mathbf{v} \cdot \mathbf{g} \right) _{i,j,k} & = & -
\frac{ \left( \rho v _x \phi \right) _{i+1/2,j,k} - \left( \rho v _x \phi  \right) _{i-1/2,j,k}}{\Delta x} + \phi _{i,j,k} 
\frac{ \left( \rho v _x \right) _{i+1/2,j,k} - \left( \rho v _x \right) _{i-1/2,j,k}}{\Delta x} \nonumber \\
& \; & - \frac{ \left( \rho v _y \phi \right) _{i,j+1/2,k} - \left( \rho v _y \phi  \right) _{i,j-1/2,k}}{\Delta y} 
+ \phi _{i,j,k} 
\frac{ \left( \rho v _y \right) _{i,j+1/2,k} - \left( \rho v _y \right) _{i,j-1/2,k}}{\Delta y} 
\nonumber \\
& \; & - \frac{ \left( \rho v _z \phi \right) _{i,j,k+1/2} - \left( \rho v _z \phi  \right) _{i,j,k-1/2}}{\Delta z} 
+ \phi _{i,j,k} \frac{\left(\rho v _z \right) _{i,j,k+1/2} - \left( \rho v _z \right) _{i,j,k-1/2}}{\Delta z} , 
\label{eq:energy_src_prior_gsub} 
\end{eqnarray}
where the temporal change in the density is evaluated from mass conservation.  Substituting
\begin{eqnarray}
\phi _{i+1/2,j,k} & = & \phi _{i,j,k} - \frac{\Delta x}{2} \mathrm{g} _{x,i+1/2,j,k} 
= \frac{\phi _{i+1,j,k} + \phi _{i,j,k}}{2}, \\
\phi _{i,j+1/2,k} & = & \phi _{i,j,k} - \frac{\Delta y}{2} \mathrm{g} _{y,i,j+1/2,k} 
= \frac{\phi _{i,j+1,k} + \phi _{i,j,k}}{2}, \\
\phi _{i,j,k+1/2} & = & \phi _{i,j,k} - \frac{\Delta z}{2} \mathrm{g} _{z,i,j,k+1/2}
= \frac{\phi _{i,j,k+1} + \phi _{i,j,k}}{2}, 
\end{eqnarray}
into Equation (\ref{eq:energy_src_prior_gsub}), we arrive at the fully conservative, discretized energy source term
\begin{eqnarray}
\left( \rho \mathbf{v} \cdot \mathbf{g} \right) _{i,j,k} = \frac{1}{2} 
\left[
\begin{aligned}
{} 
&  \hphantom{\: +} \left( \rho v _x \right) _{i+1/2,j,k} \mathrm{g} _{x,i+1/2,j,k} + \left( \rho v _x \right) _{i-1/2,j,k} \mathrm{g} _{x,i-1/2,j,k} \\ &
+ \left( \rho v _y \right) _{i,j+1/2,k} \mathrm{g} _{y,i,j+1/2,k} 
+ \left( \rho v _y \right) _{i,j-1/2,k} \mathrm{g} _{y,i,j-1/2,k} \\ &
+  \left( \rho v _z \right) _{i,j,k+1/2} \mathrm{g} _{z,i,j,k+1/2} + \left( \rho v _z \right) _{i,j,k-1/2} \mathrm{g} _{z,i,j,k-1/2}
\end{aligned} \right].
\label{eq:proper_energy_discretized}
\end{eqnarray}
There are two additional, important requirements on the energy source term: (1) the mass flux $\rho \mathbf{v}$ must be the same mass flux used in evolving the continuity equation so that the energy source term is consistent with mass conservation \citep[see][]{mikami+2008}, (2) the gravity $\mathbf{g}$ must be the average over the numerical time step.  The first requirement means that the mass flux appearing in Equation (\ref{eq:proper_energy_discretized}) should be the Riemann mass flux $\mathcal{F_\rho}$.  The second requirement arises from Equation (\ref{eq:proper2avg}), i.e, the relation
\begin{equation}
-\frac{1}{8 \pi G} \left( \frac{ \mathbf{g}(t_0 + \Delta t)^2 - \mathbf{g}(t_0)^2  }{\Delta t} \right) = -\frac{1}{4 \pi G} \mathbf{g} \cdot \frac{\mathbf{g} (t_0 + \Delta t) - \mathbf{g} (t_0)}{\Delta t}
\end{equation}
only holds when 
\begin{equation}
    \mathbf{g} = \frac{1}{2} \left[ \mathbf{g}(t_0) + \mathbf{g}(t_0+\Delta t) \right]
\end{equation}
where $t_0$ and $t_0+\Delta t$ are the times at the beginning and end of the numerical time-step $\Delta t$, respectively.  

\section{Algorithm, Properties, and Advantages} \label{sec:algorithm_properties_advantages}
\subsection{Algorithm} \label{sec:algorithm}
The fully conservative source terms in \S\ref{sec:proper_src} can be easily implemented alongside a variety of temporal integrators.  In this section, we restrict our description of the algorithm implementation to the second order accurate van-Leer predictor-corrector time integrator \citep[VL2,][]{stone+2008}, however, in Appendix \ref{sec:ext_to_rk_type_integrators}, we show how the fully conservative source terms can be extended to the strong-stability-preserving, low-storage Runge-Kutta RK2 and RK3 integrators \citep{gottlieb+2009,ketcheson2010}. 

Consider a single integration cycle of the VL2 integrator that advances cell-centered conservative variables, 
\begin{equation} \label{eq:cons}
    \mathbf{U}_{i,j,k} = \begin{bmatrix}
                    \rho \\ 
                    \rho v_x \\
                    \rho v_y \\
                    \rho v_z \\
                    E
                 \end{bmatrix},
\end{equation}
from time $t=t_0$ to time $t=t_0 + \Delta t$. Conservative variables at the initial stage, intermediate stage, and final stage are denoted $\mathbf{U}^{(0)}$, $\mathbf{U}^{(1)}$, and $\mathbf{U}^{(2)}$, respectively.  The algorithm, as presented below, assumes that the gravitational potential $\phi^{(0)}$ has already been computed from $\rho^{(0)}$ prior to executing step (1).  It then proceeds as follows:
\begin{enumerate}
    \item Advance $\mathbf{U}^{(0)}$ to $\mathbf{U}^{(1)}$ by evolving $\mathbf{U}^{(0)}$ forward in time by half a time-step, 
    \begin{equation}
        \mathbf{U}^{(1) \dagger} = \mathbf{U}^{(0)} - \frac{\Delta t}{2} \nabla \cdot \bm{\mathcal{F }} \left[\mathbf{U}^{(0)}\right], 
    \end{equation}
    where $\bm{\mathcal{F}} \left[\mathbf{U}^{(0)} \right]$ corresponds to Riemann fluxes defined at cell faces and computed from reconstructed $\mathbf{U}^{(0)}$ and $\Delta t = \Delta t (\mathbf{U}^{(0)})$ is the time-step.
    
    \item Solve for the gravitational potential $\phi^{(1)}$ (and gravity $\mathbf{g}^{(1)}$) associated with the density field $\rho^{(1)}$.  
    
    \item Apply the conservative source terms $\mathbf{S}^{(1)}$ to $\mathbf{U}^{(1)}$ following Equations (\ref{eq:proper_momentum_discretized_x}-{\ref{eq:proper_momentum_discretized_z}}) and (\ref{eq:proper_energy_discretized}),
    \begin{equation}
        \mathbf{U}^{(1)} = \mathbf{U}^{(1) \dagger} + \frac{\Delta t}{2} \mathbf{S}^{(1)},
    \end{equation}
    with
    \begin{equation}
    \mathbf{S}^{(1)}_{i,j,k} = \frac{1}{2} \begin{bmatrix}
                    0 \\ 
                    \rho^{(0)}_{i,j,k} \left( \mathrm{g}^{(0)} _{x,i+1/2,j,k} + \mathrm{g}^{(0)} _{x,i-1/2,j,k} \right)  \\
                    \rho^{(0)}_{i,j,k} \left( \mathrm{g}^{(0)} _{y,i,j+1/2,k} + \mathrm{g}^{(0)} _{y,i,j-1/2,k} \right) \\
                    \rho^{(0)}_{i,j,k} \left( \mathrm{g}^{(0)} _{z,i,j,k+1/2} + \mathrm{g}^{(0)} _{z,i,j,k-1/2} \right) \\
                    \left(
                    \begin{aligned}
                    {} 
                    & \hphantom{\; +} \mathcal{F_\rho} [\mathbf{U}^{(0)} ] _{x,i+1/2,j,k} \; \overline{\mathrm{g}} _{x,i+1/2,j,k}^{(0,1)} + \mathcal{F_\rho} [\mathbf{U}^{(0)}] _{x,i-1/2,j,k} \; \overline{\mathrm{g}} _{x,i-1/2,j,k}^{(0,1)} \\ &
                    + \mathcal{F_\rho} [\mathbf{U}^{(0)} ] _{y,i,j+1/2,k} \; \overline{\mathrm{g}} _{y,i,j+1/2,k}^{(0,1)} 
                    + \mathcal{F_\rho} [\mathbf{U}^{(0)} ] _{y,i,j-1/2,k} \; \overline{\mathrm{g}} _{y,i,j-1/2,k}^{(0,1)} \\ &
                    +  \mathcal{F_\rho} [\mathbf{U}^{(0)} ] _{z,i,j,k+1/2} \; \overline{\mathrm{g}} _{z,i,j,k+1/2}^{(0,1)} + \mathcal{F_\rho} [\mathbf{U}^{(0)} ] _{z,i,j,k-1/2} \; \overline{\mathrm{g}} _{z,i,j,k-1/2}^{(0,1)}
                    \end{aligned} \right)
                 \end{bmatrix}
    \end{equation}  
    where
    \begin{equation}
        \mathbf{\overline{g}}^{(0,1)} = \frac{1}{2} \left[ \mathbf{g}^{(0)} + \mathbf{g}^{(1)} \right],
    \end{equation}
    and $\mathcal{F}_\rho$ is the Riemann mass flux.  
    \item Advance $\mathbf{U}^{(0)}$ to $\mathbf{U}^{(2)}$ by evolving $\mathbf{U}^{(0)}$ forward in time by a full time-step,
    \begin{equation}
        \mathbf{U}^{(2) \dagger} = \mathbf{U}^{(0)} - \Delta t \nabla \cdot \bm{\mathcal{F}} \left[\mathbf{U}^{(1)} \right], 
    \end{equation}
    where $\bm{\mathcal{F}} \left[\mathbf{U}^{(1)} \right]$ corresponds to Riemann fluxes defined at cell faces and computed from reconstructed $\mathbf{U}^{(1)}$. 
    
    \item Solve for the gravitational potential $\phi^{(2)}$ (and gravity $\mathbf{g}^{(2)}$) associated with the density field $\rho^{(2)}$.  
    
    \item Apply the conservative source terms $\mathbf{S}^{(2)}$ to $\mathbf{U}^{(2)}$ following Equations (\ref{eq:proper_momentum_discretized_x}-{\ref{eq:proper_momentum_discretized_z}}) and (\ref{eq:proper_energy_discretized}),
    \begin{equation}
        \mathbf{U}^{(2)} = \mathbf{U}^{(2) \dagger} + \Delta t \mathbf{S}^{(2)},
    \end{equation}
    where
    \begin{equation}
    \mathbf{S^{(2)}}_{i,j,k} = \frac{1}{2} \begin{bmatrix}
                    0 \\ 
                    \rho^{(1)}_{i,j,k} \left( \mathrm{g}^{(1)} _{x,i+1/2,j,k} + \mathrm{g}^{(1)} _{x,i-1/2,j,k} \right)  \\
                    \rho^{(1)}_{i,j,k} \left( \mathrm{g}^{(1)} _{y,i,j+1/2,k} + \mathrm{g}^{(1)} _{y,i,j-1/2,k} \right) \\
                    \rho^{(1)}_{i,j,k} \left( \mathrm{g}^{(1)} _{z,i,j,k+1/2} + \mathrm{g}^{(1)} _{z,i,j,k-1/2} \right) \\
                    \left(
                    \begin{aligned}
                    {} 
                    & \hphantom{\; +} \mathcal{F_\rho} [\mathbf{U}^{(1)} ] _{x,i+1/2,j,k} \; \overline{\mathrm{g}} _{x,i+1/2,j,k}^{(0,2)} + \mathcal{F_\rho} [\mathbf{U}^{(1)}] _{x,i-1/2,j,k} \; \overline{\mathrm{g}} _{x,i-1/2,j,k}^{(0,2)} \\ &
                    + \mathcal{F_\rho} [\mathbf{U}^{(1)} ] _{y,i,j+1/2,k} \; \overline{\mathrm{g}} _{y,i,j+1/2,k}^{(0,2)} 
                    + \mathcal{F_\rho} [\mathbf{U}^{(1)} ] _{y,i,j-1/2,k} \; \overline{\mathrm{g}} _{y,i,j-1/2,k}^{(0,2)} \\ &
                    +  \mathcal{F_\rho} [\mathbf{U}^{(1)} ] _{z,i,j,k+1/2} \; \overline{\mathrm{g}} _{z,i,j,k+1/2}^{(0,2)} + \mathcal{F_\rho} [\mathbf{U}^{(1)} ] _{z,i,j,k-1/2} \; \overline{\mathrm{g}} _{z,i,j,k-1/2}^{(0,2)}
                    \end{aligned} \right)
                 \end{bmatrix}
    \end{equation}  
    and
    \begin{equation}
        \mathbf{\overline{g}}^{(0,2)} = \frac{1}{2} \left[ \mathbf{g}^{(0)} + \mathbf{g}^{(2)} \right].
    \end{equation}
    
    \item Replace $t_0 + \Delta t \rightarrow t_0$, $\phi^{(2)} \rightarrow \phi^{(0)}$, and $\mathbf{U}^{(2)} \rightarrow \mathbf{U}^{(0)}$.
    
\end{enumerate}
The algorithm is of second-order spatial and temporal accuracy and only requires two Poisson solves per time-step (steps 2 and 5).  We assumed that $\phi^{(0)}$ was computed prior to executing step (1), however, this requirement only manifests itself in the very first cycle of the numerical integration; all subsequent cycles are supplied $\phi^{(0)}$ from steps (5) and (7).  The algorithm requires that the fluxes $\mathcal{F_\rho} \left[\mathbf{U}^{(0)} \right]$ and $\mathcal{F_\rho} \left[\mathbf{U}^{(1)} \right]$ are the same mass fluxes applied in steps (1) and (4), respectively; we note that these fluxes can be computed via any Riemann solver or reconstruction method.  The energy source terms $S^{(\ell)}_{E,i,j,k}$ in steps (5) and (6) are dependent on the gravity $\mathbf{g}^{(0)}$ and $\mathbf{g}^{(\ell)}$, hence, the fully conservative algorithm requires (a) additional memory to store the gravitational potential at the initial stage $\phi^{(0)}$ and (b) that the continuity equation be evolved prior to the application of the energy source term such that the density at the advanced stage $\rho^{(\ell)}$ can provide the gravity at the advanced stage $\mathbf{g}^{(\ell)}$ (i.e., steps 1-2 and steps 4-5).

\subsection{Total Momentum and Total Energy Conservation} \label{sec:conservation}
The algorithms described in \S\ref{sec:algorithm} and Appendices \S\ref{sec:rk2} and \S\ref{sec:rk3} will guarantee total momentum and total energy conservation since the solutions satisfy Equation (\ref{eq:proper1}) and Equations (\ref{eq:proper2}). 
However, in order to achieve round-off error conservation using source terms, the solution to the discretized Poisson equation (Equation \ref{eq:poisson_discretized}) must be accurate to round-off error.  FFT-based Poisson solvers \citep[e.g.,][]{hockney+1988,moon+2019} are capable of producing such machine-accurate solutions.  In contrast, multigrid methods \citep[e.g.,][Tomida et al., in preparation]{matsumoto+2003,ricker2008} often yield gravitational potential solutions that contain residual errors, thus, total momentum and total energy may not be conserved to round-off error.  An interesting property of the fully conservative source term algorithm in \S\ref{sec:algorithm} (and Appendices \ref{sec:rk2} and \ref{sec:rk3}) is that the gravity at intermediate stage(s) does not enter the energy source term when advancing conservative variables from $t_0$ to $t_0+ \Delta t$ in the final stage.  Thus, the algorithm guarantees total energy conservation if the gravity at the initial and final stages are accurate to round-off error, even if the intermediate gravity contains a residual error. The momentum source terms (Equations \ref{eq:proper_momentum_discretized_x}-\ref{eq:proper_momentum_discretized_z}) only conserve total linear momentum to round-off error when the difference between the implied density from the discretized Poisson equation (Equation \ref{eq:poisson_discretized}) is equivalent to the density $\rho_{i,j,k}$, i.e., $\Delta \rho_{i,j,k} = 0$, where
\begin{eqnarray}
 \Delta \rho_{i,j,k} 
  = \frac{1}{4\pi G} & &
 \left[\frac{\phi  _{i+1,j,k} - 2 \phi _{i,j,k} + \phi _{i-1,j,k}}{\Delta x ^2}
 +  \frac{\phi _{i,j+1,k} - 2 \phi  _{i,j,k} + \phi _{i,j-1,k}}{\Delta y ^2} \right. \nonumber \\
 & & \left.
 + \frac{\phi  _{i,j,k+1} - 2 \phi  _{i,j,k} + \phi _{i,j,k-1}}{\Delta z ^2} \right] - \rho_{i,j,k}.  
\end{eqnarray}
However, we note that even in the presence of a residual error, we can still guarantee round-off error linear momentum conservation using the source term scheme by adding a uniform, constant ``corrective acceleration" $\mathbf{g_\mathrm{corr}}$ to the gravity $\mathbf{g}$ appearing in Equations (\ref{eq:proper_momentum_discretized_x}-\ref{eq:proper_momentum_discretized_z}), where 
\begin{equation}
    \mathbf{g_\mathrm{corr}} = \frac{\sum_{i,j,k} \Delta \rho_{i,j,k} \mathbf{g}_{i,j,k}}{\sum_{i,j,k} \rho_{i,j,k}}.
\end{equation}
This corrective gravity will ensure linear momentum conservation since 
\begin{equation}
    \sum_{i,j,k} \rho_{i,j,k} \left( \mathbf{g}_{i,j,k} + \mathbf{g_\mathrm{corr}} \right) = 0.
\end{equation}
The corrective gravity $\mathbf{g_\mathrm{corr}}$ is uniform and will therefore not introduce a stress or tidal force.
\subsection{Suppression of Anomalous Accelerations} 
\label{sec:curl_g_zero}
A final major advantage of the fully conservative source-term-based scheme is that it guarantees that the computed gravity is curl-free, where $\nabla \times \mathbf{g}$ is defined by Equations (\ref{eq:curlgx}-\ref{eq:curlgz}).   In contrast, a common discretization of the gravitational stress tensor $\mathbf{\tilde{T}_g}$ (see Equations \ref{eq:Tg_tildexx}-\ref{eq:Tg_tildezx}) does not.  $\mathbf{\tilde{T}_g}$ represents the numerical discretization of the gravitational stress tensor as employed by \texttt{Athena++ v19.0} \citep{stone+2020} and \texttt{Athena} \citep{stone+2008}.  What are the consequences of violating the $\nabla \times \mathbf{g}=0$ constraint? We find that the gravitational stress tensor $\mathbf{\tilde{T}_g}$ can produce anomalous accelerations of such large magnitude that they can compromise simulations of self-gravitating flows.

We investigate these anomalous accelerations via an illustrative (and analytic) model problem.  Consider the gravity outside a spherical body of mass $M$ and radius $R$ surrounded by a spherically-symmetric atmosphere of mass $M_\mathrm{a}$.  Let the mass profile (for $r > R$) be expressed by
\begin{equation}
    M(r) = M + M_a \left[1 - \left( \frac{r}{R} \right)^{-2} \right].
    \label{eq:atmos_mass_profile}
\end{equation}
\noindent For $r>R$, the gravitational potential, $\phi$, is
\begin{equation} \label{eq:analytic_phi}
    \phi (r) = G \int_\infty^r \frac{M(r')}{r'^2} dr' = -\frac{G M}{r} - \frac{G M_a}{r} \left[1- \frac{1}{3} \left( \frac{r}{R} \right)^{-2} \right].
\end{equation}
\noindent Thus, the true gravitational accelerations at $r$ (for $r>R$) are, 
\begin{equation}
    \mathbf{g_\mathrm{true}} = -\nabla \phi = -\frac{G M(r)}{r^2} \hat{r},
\end{equation}
where $ \hat{r} $ denotes the unit vector in the radial direction.

Now we evaluate the effects of discretization on this model problem for uniform Cartesian grids using the same $i,j,k$ notation from \S\ref{sec:proper_src}.  Let the gravitational potential $\phi_{i,j,k}$ be set by the analytic values in Equation (\ref{eq:analytic_phi}) and let the density $\rho_{i,j,k}$ be set through the discretized Poisson equation (Equation \ref{eq:poisson_discretized}).  This prescription for $\rho_{i,j,k}$ may be inconsistent with the mass distribution $M(r)$. 

For the source-term-based approach, the discretized gravitational accelerations $\mathbf{g_\mathrm{src}}$ can be obtained from Equations (\ref{eq:proper_momentum_discretized_x}-\ref{eq:proper_momentum_discretized_z}). For the ``gravity flux" scheme, in conjunction with $\mathbf{\tilde{T}_g}$, the gravitational accelerations $\mathbf{g_\mathrm{flx}}$ are computed from the discretized divergence of the gravitational stress tensor ($-(\nabla \cdot \mathbf{\tilde{T}_g})_{i,j,k}$) divided by the density $\rho_{i,j,k}$.

For our model problem, both $\mathbf{g_\mathrm{src}}$ and $\mathbf{g_\mathrm{flx}}$ have analytic, albeit complicated, forms.  For $\Delta x = \Delta y = \Delta z = d$, we can expand each in a Taylor series in $d$.  This yields leading order terms (i.e., corresponding to $d \rightarrow 0$) that recover the true gravitational accelerations $\mathbf{g}_\mathrm{true}$, followed by error terms,
\begin{eqnarray}
    \mathbf{g_\mathrm{src}} & = & \mathbf{g_\mathrm{true}} + d^2 \bm{\varepsilon}_\mathrm{src} + \mathcal{O}(d)^4, \\
    \mathbf{g_\mathrm{flx}} & = & \mathbf{g_\mathrm{true}} + d^2 \bm{\varepsilon}_\mathrm{flx} + \mathcal{O}(d)^4.
\end{eqnarray}
Both schemes are therefore second-order accurate. The $d^2 \bm{\varepsilon}_\mathrm{src}$ and $d^2 \bm{\varepsilon}_\mathrm{flx}$ terms can introduce an error to both the magnitude and direction of the gravity for both the source-term-based and ``gravity flux" scheme.  Expanding $\lvert d^2 \bm{\varepsilon} \rvert$ for small $M_a$ at the $z=0$ plane, we obtain
\begin{equation}
    \lvert d^2 \bm{\varepsilon}_\mathrm{src} \rvert =  d^2 \left[ \frac{G M}{4 r^4} \left( \frac{17 + 15 \cos 4 \varphi }{2} \right)^{1/2} + \mathcal{O} (M_a)^1 \right],
\end{equation}
\begin{equation}
    \lvert d^2 \bm{\varepsilon}_\mathrm{flx} \rvert = d^2 \left[ \frac{3 G M^2}{32 r^2 R^2 M_a} \left(\frac{ 143 + 60 \cos 4 \varphi - 75 \cos 8 \varphi}{2}\right)^{1/2} + \mathcal{O} \left( M_a \right)^0 \right],
\end{equation}
where $\varphi$ is the spherical azimuthal angle. 

For the source-term-based scheme, the error term $\lvert d^2 \bm{\varepsilon}_\mathrm{src} \rvert$ is independent of $M_a$ in the limit of small $M_a$.  Strikingly, we find that $\lvert d^2 \bm{\varepsilon}_\mathrm{flx} \rvert $ is divergent in the limit of small $M_a$.  By equating $\lvert d^2 \bm{\varepsilon}_\mathrm{flx} \rvert$ to $\lvert \mathbf{g_\mathrm{true}} \rvert $, we identify that the critical atmosphere mass to central mass ratio that produces error terms with magnitudes of the same order of the true accelerations (in the vicinity of $r \sim R$) is
\begin{equation}\label{eq:critical}
    \left( \frac{M_a}{M} \right)_\mathrm{critical} = \frac{3}{4} \left( \frac{R}{d} \right)^{-2}
\end{equation}  
for the ``gravity flux" scheme (in conjunction with $\mathbf{\tilde{T}_g}$).  Above, the ratio $R/d$ represents the number of grid cells resolving the central body's radius.  For $R/d = 10$, Equation (\ref{eq:critical}) gives $\left( M_a/M \right)_\mathrm{crit} \simeq 10^{-2}$. 

Figure \ref{fig:error_model} shows the magnitude and direction of the radial and azimuthal components of $d^2 \bm{\varepsilon}_\mathrm{src} $ and $d^2 \bm{\varepsilon}_\mathrm{flx} $ for model parameters $G=M=R=1$, $R/d = 10$, and $M_a$/$M$ = 10$^{-2}$.  We note that for this model problem, our newly proposed discretization of the gravitational stress tensor $\mathbf{T_g}$ yields $ d^2 \bm{\varepsilon} $ errors equivalent to $d^2 \bm{\varepsilon}_\mathrm{src}$.

\section{Implementation in Athena++} \label{sec:implementation_in_athena++}
We implement the fully conservative, source-term-based numerical algorithm for self-gravitating (magneto)hydrodynamics in the \texttt{Athena++} framework for static, uniform, Cartesian meshes.  We test our implementation via three test problems that target investigating the scheme's (1) error convergence (see \S\ref{sec:algorithm}), (2) total momentum and total energy conservation (see \S\ref{sec:conservation}), (3) suppression of anomalous accelerations (see \S\ref{sec:curl_g_zero}), (4) durability against residual errors (see \S\ref{sec:conservation}), and (5) ability to maintain self-gravitating equilibria.  For each test, we employ the FFT Poisson solver included in \texttt{Athena++} \citep{hockney+1988,stone+2020}, an HLLC Riemann solver \citep{toro2009}, and a gamma-law equation of state $P=(\gamma-1)e$ where $\gamma$ is the adiabatic index.  Unless otherwise stated, all test problems apply the second-order accurate van-Leer VL2 integrator \citep{stone+2008} and piecewise linear (PLM) reconstruction.

\subsection{Spitzer Sheets (1-D Equilibria)} \label{sec:spitzer}
To test for second-order convergence and total linear momentum conservation in our scheme implementation, we first study the advection of 1-D self-gravitating equilibria \citep[i.e., Spitzer sheets, ][]{spitzer1942} on periodic meshes.  Spitzer sheet equilibria satisfy the requirement of hydrostatic equilibrium,
\begin{equation}
    -\frac{1}{\rho} \frac{d P}{d z} - \frac{d \phi}{d z} = 0, 
\end{equation}
Poisson's equation \citep[in conjunction with Jeans swindle,][appropriate for periodic boundary conditions]{jeans1902},
\begin{equation}
    \frac{d^2 \phi}{d z^2} = 4 \pi G (\rho - \overline{\rho}),
\end{equation}
and a polytropic pressure profile,
\begin{equation}
    P = K \rho^\mathrm{\Gamma},
\end{equation}
where $\rho$ is the density profile, $P$ is the pressure profile, $\phi$ is the gravitational potential, $\overline{\rho}$ is the mean density, $\Gamma$ is the polytropic index, and $K$ is a constant that sets the specific entropy of the sheet.

\begin{figure}
    \centering
    \includegraphics[width=\textwidth]{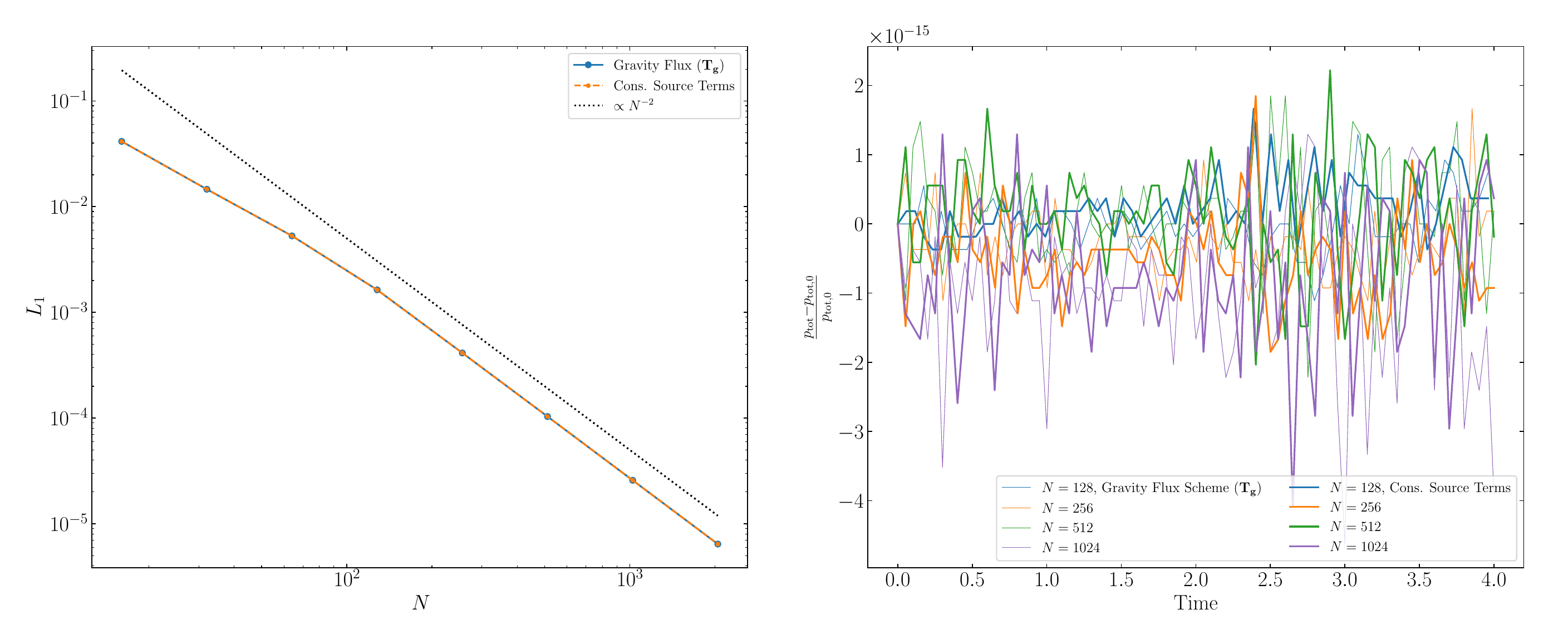}
    \caption{(Left): $L_1$ error convergence for the test problem of advecting a Spitzer sheet once across a periodic domain using momentum ``gravity fluxes" with $\mathbf{T_g}$ (blue) and fully conservative momentum source terms (orange) in conjunction with a VL2 integrator and PLM reconstruction.  Second order error convergence is shown in dotted black.  (Right): Conservation of total momentum $p_\mathrm{tot}$ as a function of time for the ``gravity flux" scheme (fine) and source term scheme (bold) at several numerical resolutions $N$.}
    \label{fig:spitzer}
\end{figure}

We choose parameters $G = K = 1.0$, $\Gamma = \gamma = 1.2$ (yielding isentropic equilibria), $\overline{\rho} = 0.3$, and mesh size $L_z = 4$.  The simulations are initialized with round-off error accurate solutions to conservative variables at cell centers (i.e., Equation \ref{eq:cons}) for 1-D meshes resolved by $N=(16,32,64,128,256,512,1024,2048)$ cells.  For our model parameters, the equilibrium solutions have density contrast $\rho_\mathrm{max}/\rho_\mathrm{min} \sim 10^4$, where $\rho_\mathrm{max}$ and $\rho_\mathrm{min}$ are the maximum and minimum densities in the sheet. We advect the 1-D equilibria at a velocity $v_z = 1$, thus requiring an integration from $t=t_i=0$ to $t=t_f=4$ (in code units) for a full, single translation of the sheet across the periodic domain.   After advection, we measure the $L_1$ error,
\begin{equation}
L_1 = \sum_{i} \lvert \rho_{i} - \rho_{\mathrm{exact}} \rvert \Delta z.
\end{equation}
Figure \ref{fig:spitzer} (left) presents the $L_1$ error convergence analysis, for schemes employing (a) fully conservative momentum source terms and (b) momentum ``gravity fluxes" (in conjunction with $\mathbf{T_g}$).  Both (a) and (b) use the conservative energy source term in Equation (\ref{eq:proper_energy_discretized}).  Both schemes converge at second order (i.e., $L_1 \propto N^{-2}$).  Figure \ref{fig:spitzer} (right) gives the time evolution of the total momentum $p_\mathrm{tot}$; both schemes conserve $p_\mathrm{tot}$ to round-off error.
\subsection{Jeans Linear Waves in 3-D} 
\label{sec:jeans}
Next, we study Jeans linear waves in 3-D: a plane wave perturbation aligned with the $x_1$-axis with wavelength $\lambda$ (and perturbed density $\delta \rho$, perturbed pressure $\delta P$, and perturbed velocity along the $x_1$-axis, $\delta v_1$),  
\begin{equation}
     \delta \rho \left(x_1 \right) = \rho_0 A \sin \left(\frac{2 \pi}{\lambda} x_1 \right)
\end{equation}
\begin{equation}
    \delta P (x_1) = \gamma P_0 A \sin \left( \frac{2 \pi}{\lambda} x_1 \right)
\end{equation}
\begin{equation}
    \delta v_1 (x_1) =
    \begin{dcases}
    \hphantom{-} \frac{\sqrt{ \lvert \omega^2 \rvert }}{2 \pi / \lambda} A \cos \left( \frac{2 \pi}{\lambda} x_1 \right), & \mathrm{if} \; \omega^2  < 0 \\
    - \frac{\sqrt{ \lvert \omega^2 \rvert }}{2 \pi / \lambda} A \sin \left( \frac{2 \pi}{\lambda} x_1 \right), & \mathrm{if} \;  \omega^2  > 0
    \end{dcases}
\end{equation}
is added atop an otherwise static uniform background with density and pressure, $\rho_0$ and $P_0$.  Above, $\gamma$ is the adiabatic index, $A$ is the wave amplitude, and 
\begin{equation}
    \omega^2 = \left(\frac{2 \pi}{\lambda}\right)^2 c_{s,0}^2 - 4 \pi G \rho_0
\end{equation}
is the Jeans dispersion relation, where $c_{s,0}^2 = \gamma P_0/ \rho_0$ is the adiabatic sound speed squared  and $G$ is the graviational constant.  Setting $\omega^2=0$ in the dispersion relation gives the Jeans wavelength, 
\begin{equation}
   \lambda_J = \sqrt{ \pi c_s^2 / G \rho }.
\end{equation}
\noindent When $\lambda/\lambda_J > 1$, $\omega^2 < 0$ and the plane wave perturbation is unstable to gravitational collapse \citep[i.e., the Jeans instability,][]{jeans1902}.   When $\lambda/\lambda_J < 1$, $\omega^2 > 0$ and the plane wave perturbation yields a stable propagating wave with oscillation period $2 \pi / \omega$.

We rotate the wavevector of the plane wave perturbation (via a coordinate transformation) so that it is not parallel to any grid axis.  Our choices for mesh size and rotation angles are adopted from \cite{gardiner+2008} and \cite{stone+2008} and guarantee that (1) the wavevector does not lie along a cell diagonal, (2) the plane wave has a perturbation wavelength $\lambda=1$, and (3) there is one wave period along each grid direction.  In our new coordinate system, the mesh has $(L_x, L_y, L_z) = (3, 3/2, 3/2)$ resolved by $(2N, N, N)$ cells and the wavevector of the plane wave perturbation is $\mathbf{k} = \left[k_x, k_y, k_z \right] = 2 \pi/\lambda \left[1/3, 2/3, 2/3 \right]$. The boundary conditions are periodic. We set $\rho_0=1, P_0 = 1/\gamma$, $\gamma=5/3$, and $A=10^{-6}$.  Simulations are initialized by setting cell-centered conservative variables to their analytic values. 

\subsubsection{Stable ($\omega^2 > 0$)}
\begin{figure}
    \centering
    \includegraphics[width=\textwidth]{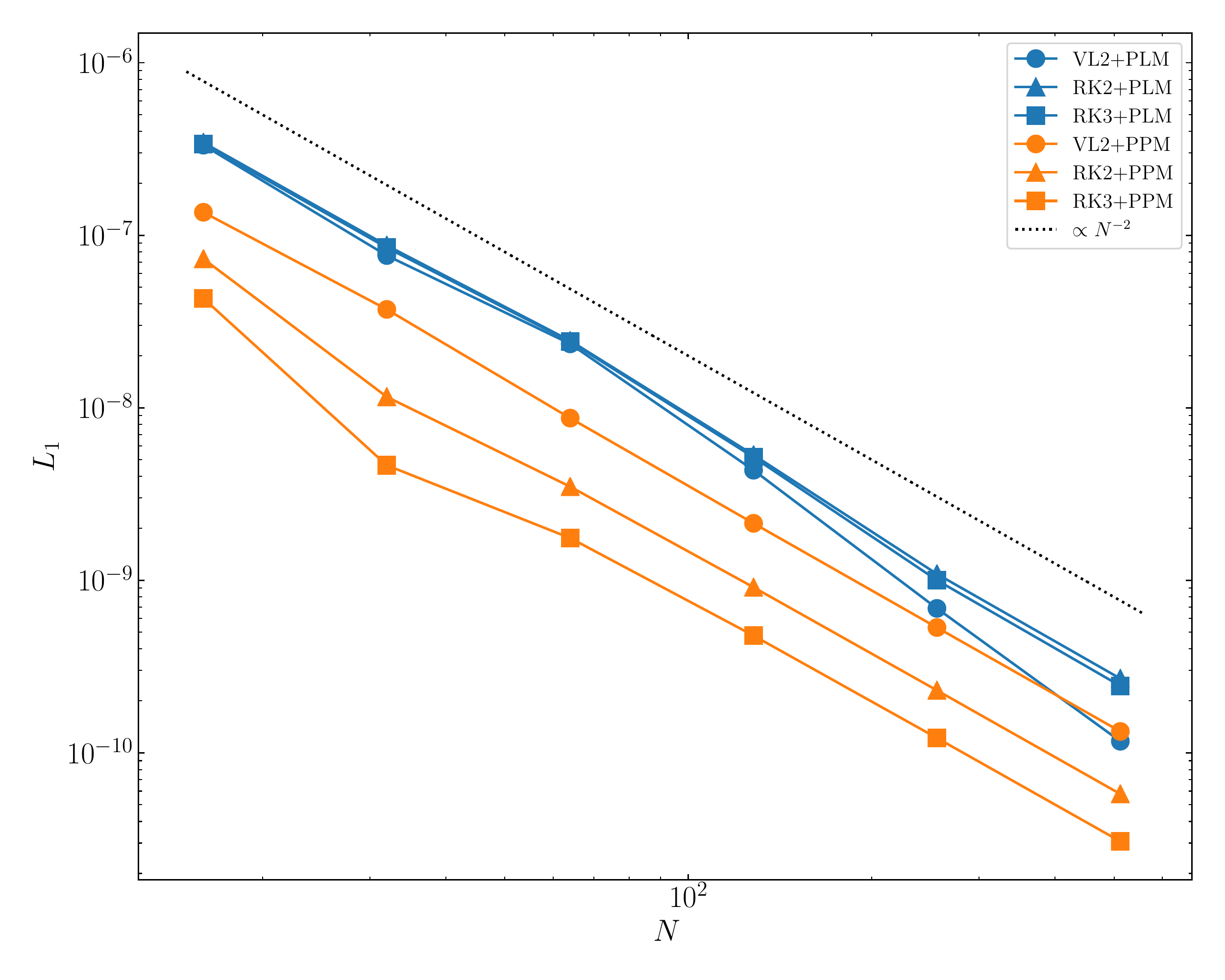}
    \caption{$L_1$ error convergence for the test problem of propagating stable Jeans linear waves (with $\lambda/\lambda_J=1/2$) across a 3-D periodic mesh for a single period ($2 \pi / \omega$) using the fully conservative source term scheme in conjunction with all combinations of temporal integrators (VL2: circles, RK2: triangles, RK3: squares) and reconstruction methods (PLM: blue, PPM: orange).}
    \label{fig:jeans_convergence}
\end{figure}

We first investigate Jeans stable linear waves with $\lambda/\lambda_J = 1/2$. We resolve the 3-D meshes with a varying number of cells, i.e., $N=(8,16,32,64,128,256)$.  After evolving the stable wave perturbation described above for a single oscillation period $2 \pi/\omega$, we measure the $L_1$ error (now modified in 3-D) 
\begin{equation}
L_1 = \frac{ \sum_{i,j,k} \lvert \rho_{i,j,k} - \rho_{\mathrm{exact}} \rvert \Delta x \Delta y \Delta z}{\sum_{i,j,k} \Delta x \Delta y \Delta z}. 
\end{equation}
For this analysis, we consider all temporal integrators described in this work: (1) the second-order accurate van Leer integrator (VL2), (2) the second-order accurate Runge-Kutta integrator (RK2), and (3) and the third-order accurate Runge-Kutta integrator (RK3).  We also study two different reconstruction methods: (1) piecewise linear reconstruction (PLM) and (2) piecewise parabolic reconstruction (PPM).  Despite the use of higher order temporal integrators and reconstruction methods, the scheme is limited to second-order accuracy due to our evaluation of the gravity. Figure \ref{fig:jeans_convergence} demonstrates that strict second-order error convergence is observed for all integrator/reconstruction combinations.  Almost universally, higher order temporal integrators and reconstruction methods lower the $L_1$ error for a given $N$.

\subsubsection{Unstable ($\omega^2 < 0$)}
\begin{figure}
    \centering
    \includegraphics[width=\textwidth]{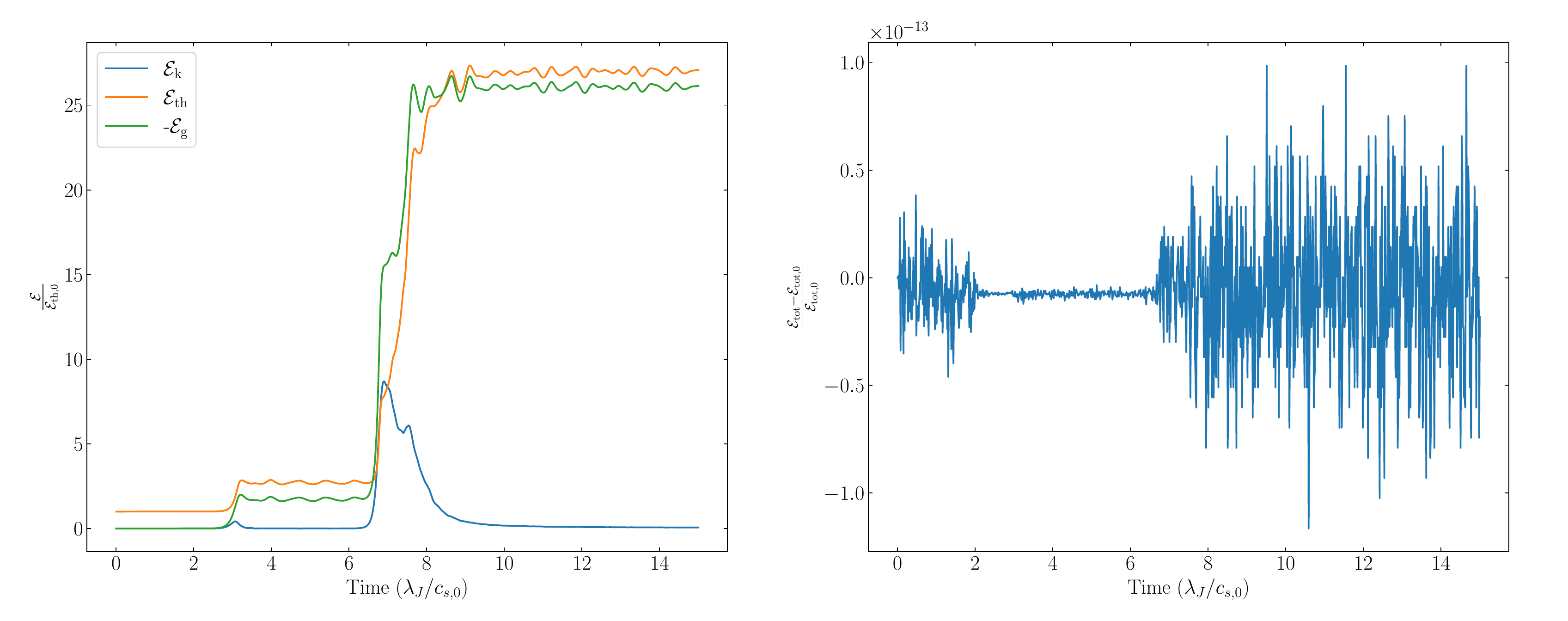}
    \caption{(Left): Components of the volume-integrated energy (kinetic: blue, thermal: orange, gravitational: green) as a function of time in the Jeans instability problem (with $\lambda/\lambda_J=3/2$) when applying the fully conservative source-term based scheme.  As in \cite{hanawa2019}, the volume-integrated energies are normalized by the initial total thermal energy of the mesh, $\mathcal{E}_\mathrm{th,0}$.  (Right): Conservation to total energy $\mathcal{E}_\mathrm{tot}$ as a function of time.  }
    \label{fig:jeans_energy}
\end{figure}

We now turn to the unstable case.  Figure \ref{fig:jeans_energy} left tracks each component of the energy for the evolution of a perturbation with $\lambda/\lambda_J = 3/2$.  The Jeans instability test has volume-integrated kinetic energy ($\mathcal{E}_\mathrm{k}$), volume-integrated thermal energy ($\mathcal{E}_\mathrm{th}$), and volume-integrated gravitational energy ($\mathcal{E}_\mathrm{g}$) varying substantially over the course of the unstable evolution.  The plane wave perturbation first collapses into a sheet at $t\sim3 \; \lambda_J/c_{s,0}$. At $t\sim7 \; \lambda_J/c_{s,0}$, the sheet collapses into filaments.  Figure \ref{fig:jeans_energy} right plots the total energy, 
\begin{equation} \label{eq:jeans_total_energy}
\mathcal{E}_\mathrm{tot} = \sum_{i,j,k} \left[ \frac{1}{2} \rho_{i,j,k} \lvert \mathbf{v}_{i,j,k} \rvert^2 + \frac{P_{i,j,k}}{\gamma-1} + \frac{1}{2} \rho_{i,j,k} \phi_{i,j,k} \right] \Delta x \Delta y \Delta z,
\end{equation}
\noindent for the duration of the integration ($t \sim10 \lambda_J/c_{s,0}$).  Note that in Equation (\ref{eq:jeans_total_energy}),  we evaluate the total gravitational energy following the prescription set forth in Equation (\ref{eq:eg_sum_discretized}), but recall that the two forms of the total gravitational energy are equivalent via Equation (\ref{eq:volume_int_equal}).  The fully conservative source term scheme conserves total energy to round-off error, despite the large changes in the magnitude of each volume-integrated energy component.  We also confirm that total linear momentum is conserved throughout the evolution.

\subsection{Polytropes (3-D Equilibria)} \label{sec:polytropes}
The next test problems evolve 3-D equilibria of self-gravitating polytropes.  We discretize the analytic model presented in section \S\ref{sec:curl_g_zero}.  The central body of mass $M$ and radius $R$ is modeled as a $\Gamma = \gamma = 2$ polytrope. Such equilibria have analytic solutions to the Lane-Emden equation,
\begin{equation}
\rho (r) = \rho_c \frac{\sin(\alpha r)}{\alpha r},
\end{equation}
\noindent where
\begin{equation}
\alpha = \sqrt{\frac{2 \pi G}{p_c}} \rho_c,
\end{equation}
\noindent and $\rho_c$ and $p_c$ are the central density and pressure of the polytrope, respectively. 

\begin{figure}
    \centering
    \includegraphics[width=\textwidth]{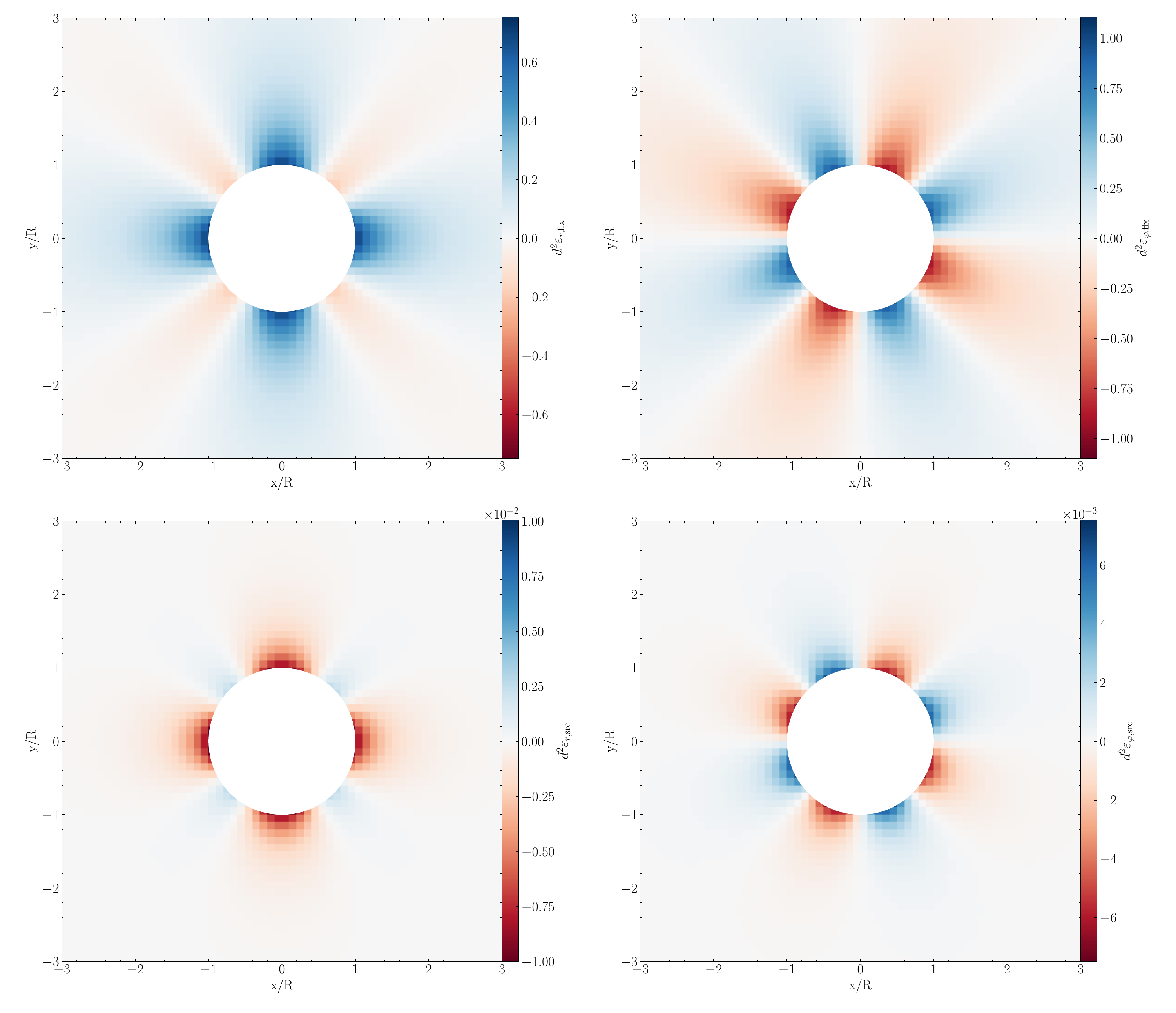}
    \caption{Radial (left) and azimuthal (right) components of the error terms $d^2 \bm{\varepsilon}$ associated with (top) the momentum ``gravity flux" scheme with $\mathbf{\tilde{T}_g}$ and (bottom) the conservative momentum source term, for a model problem described in the main text (with model parameters $G=M=R=1$, $R/d=10$, and $M_a/M = 10^{-2}$) at the $z=0$ plane.  For the radial contributions (left), red and blue correspond to inwardly and outwardly directed accelerations, respectively.  For the azimuthal contributions (right), red and blue correspond to clockwise and counterclockwise accelerations, respectively.  The colorbars are not identical for all panels; the maximum and minimum of the colorbars for the $d^2 \bm{\varepsilon}_\mathrm{src}$ panels were reduced by nearly two orders of magnitude compared to $d^2 \bm{\varepsilon}_\mathrm{flx}$ panels, so that features could be visible.}
    \label{fig:error_model}
\end{figure}

For $r>R$, we shift from the polytropic profile to the atmospheric density profile consistent with Equation (\ref{eq:atmos_mass_profile}).  We choose an atmosphere mass of $M_a = 10^{-2} M$, as in Figure \ref{fig:error_model}.  A contact discontinuity exists at $r=R$ in the initial condition of our planet/atmosphere density profile, however, pressure is continuous everywhere.  We select a pressure profile for $r>R$ that guarantees the atmosphere is in hydrostatic equilibrium.

We set $G=M=R=1$.  The 3-D mesh is uniform and Cartesian, with $(L_x, L_y, L_z) = (8R, 8R, 8R)$ resolved by $N=80^3$ cells, i.e, $R/d \sim 10$ (as in Figure \ref{fig:error_model}).  We impose periodic boundary conditions, therefore, the analytic equilibrium slightly differs from the numerical equilibrium.  The dynamical time of the polytrope is $\tau = R/ v_\mathrm{esc}$, where $v_\mathrm{esc}$ is the escape velocity, $v_\mathrm{esc} = (2 G M / R)^{1/2}$.

\subsubsection{Anomalous Accelerations}
\label{sec:suppression}
\begin{figure}
    \centering
    \includegraphics[width=\textwidth]{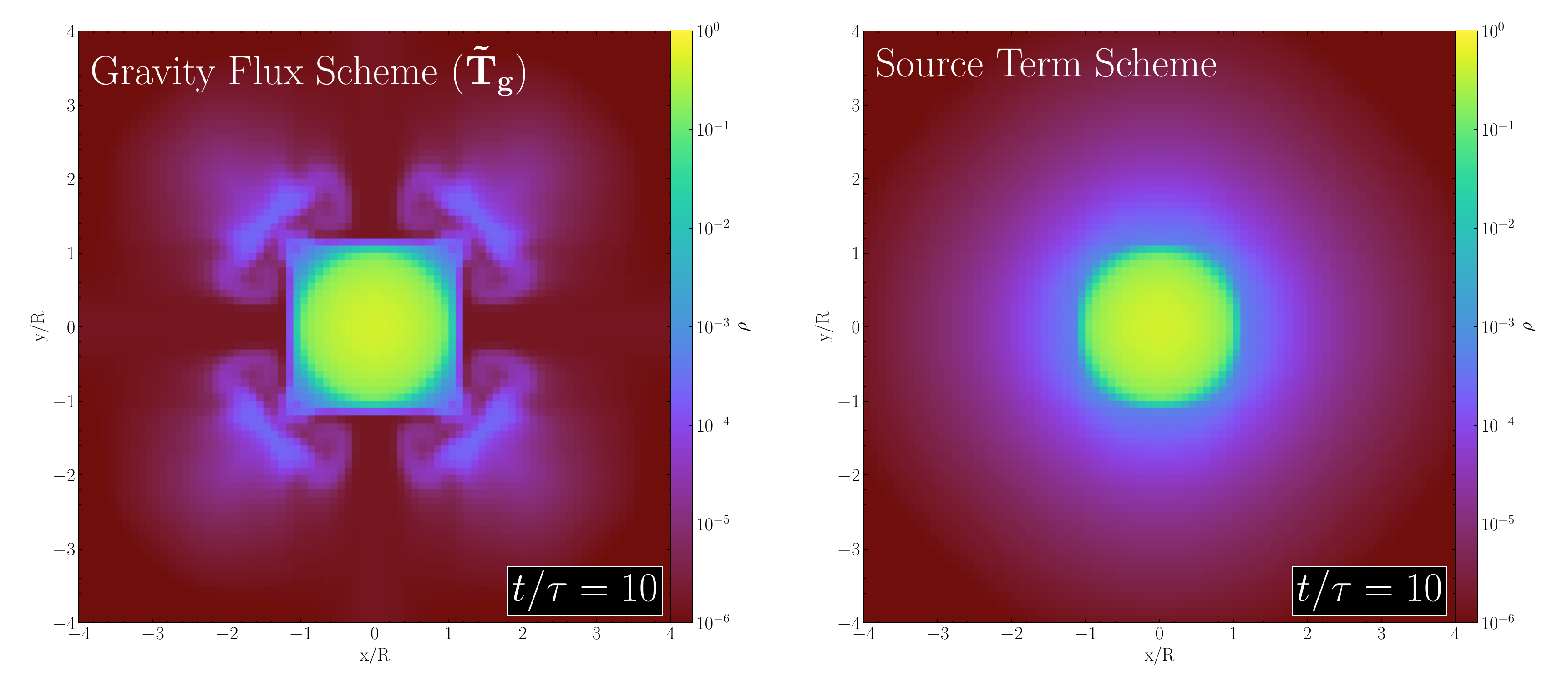}
    \caption{Equatorial density slices ($z=0$) of a 3-D polytrope with an overlying atmosphere after evolving the system for ten dynamical times ($t/\tau =10$) using (left) momentum ``gravity fluxes" (with gravitational stress tensor $\mathbf{\tilde{T}_g}$) and (right) the fully conservative momentum source terms. }
    \label{fig:polytrope}
\end{figure}

We first demonstrate that the source term scheme can maintain the hydrostatic equilibrium of a spherical body surrounded by an atmosphere for many dynamical times.  Figure \ref{fig:polytrope} presents density slices through the equators of the polytropes (and atmospheres) after integrating the equilibria for $t/\tau=10$ using (left) momentum ``gravity fluxes" with $\mathbf{\tilde{T}_g}$ (as in \texttt{Athena++ v19.0} and \texttt{Athena}) and (right) the fully conservative momentum source terms described in this work.  Both (left) and (right) use the conservative energy source term in Equation (\ref{eq:proper_energy_discretized}).

Anomalous accelerations when employing $\mathbf{\tilde{T}_g}$ destroy the equilibrium atmosphere by producing over-pressured and under-pressured regions near the surface of the polytrope (see Figure \ref{fig:error_model}); resulting pressure gradient forces yield inflow along grid axes and outflow along diagonals yielding an $m=4$ component in the aftermath.

\begin{figure}
    \centering
    \includegraphics[width=\textwidth]{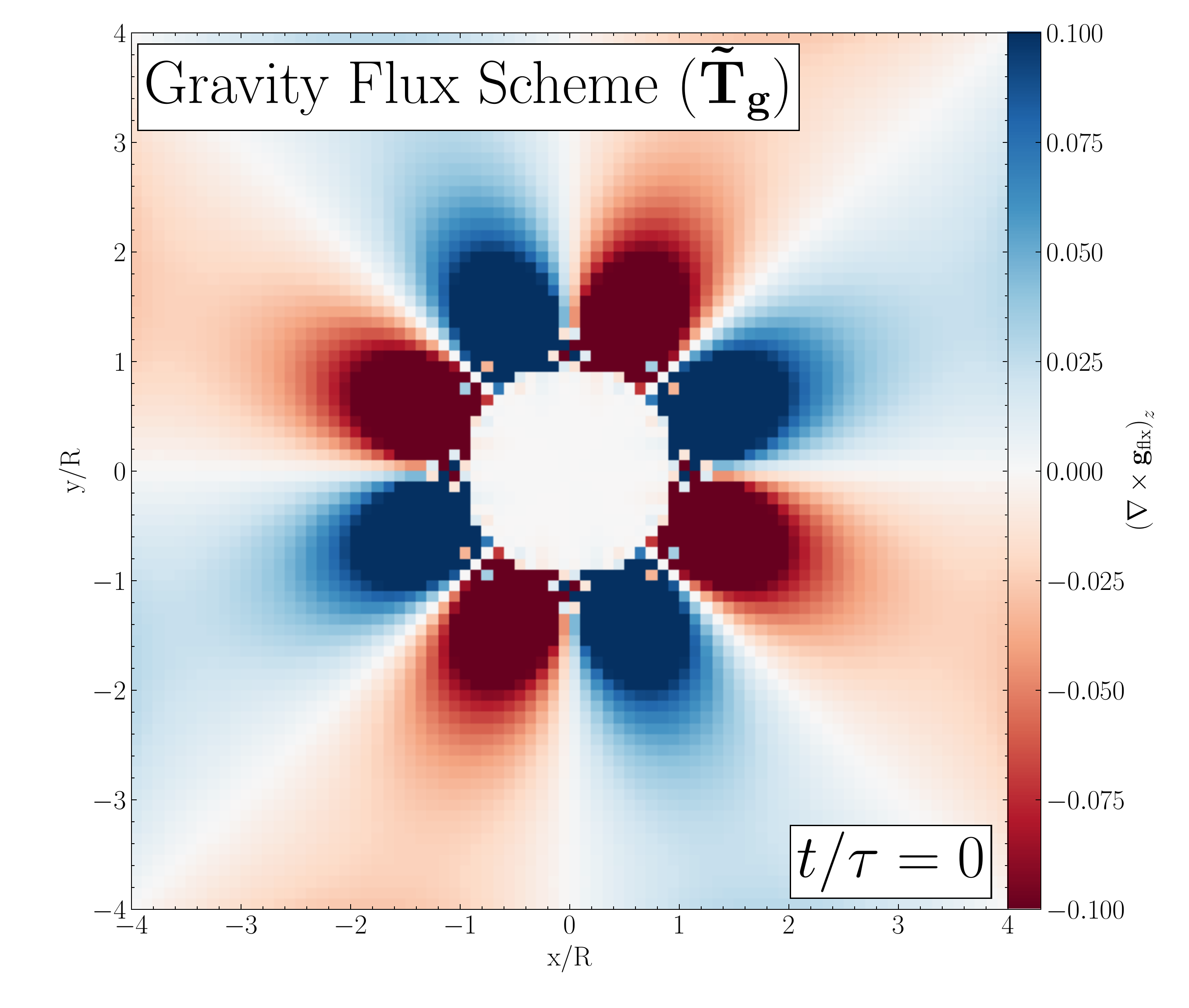}
    \caption{Equatorial slice ($z=0$) through the initial state ($t/\tau=0$) of a 3-D polytrope with an overlying atmosphere showing the $z-$component of $\nabla \times \mathbf{g_\mathrm{flx}}$, where $\mathbf{g_\mathrm{flx}}$ is the gravity obtained from the ``gravity flux" scheme in conjunction with the gravitational stress tensor $\mathbf{\tilde{T}_g}$.}
    \label{fig:curl_g}
\end{figure}

These inflows and outflows are related to the violation of the $\nabla \times \mathbf{g} = 0$ constraint when using $\mathbf{\tilde{T}_g}$. Figure \ref{fig:curl_g} shows departures from $\left(\nabla \times \mathbf{g_\mathrm{flx}} \right)_z = 0$ in the $z=0$ plane in the initial state ($t/\tau=0$), where $\mathbf{g_\mathrm{flx}}$ is the gravity obtained from the ``gravity flux" scheme in conjunction with $\mathbf{\tilde{T}_g}$ and 
$(\nabla \times \mathbf{g_\mathrm{flx}})_z$ is computed following Equation (\ref{eq:curlgz}).
The source term scheme and the ``gravity flux" scheme in conjunction with $\mathbf{T_g}$ (not $\mathbf{\tilde{T}_g}$) both guarantee that the computed gravity is curl-free to round-off error. Even after ten dynamical times, the source term scheme maintains the initial equilibrium. The ``gravity flux" scheme with $\mathbf{T_g}$ gives evolution nearly indistinguishable from the fully conservative source term scheme.  

It is worth noting that total momentum and total energy are conserved to numerical round-off error for the $\mathbf{\tilde{T}_g}$, $\mathbf{T_g}$, and conservative source term runs.  We have also confirmed that each scheme conserves total linear momentum when the 3-D polytrope and overlying atmosphere is advected across the diagonal of the mesh with velocity $\lvert \mathbf{v} \rvert = v_\mathrm{esc}$.
\subsubsection{Residual Errors}
\label{sec:residual}
Until this point, we have only considered an FFT-based Poisson solver \citep{hockney+1988,stone+2020} that produces machine-accurate solutions to the discretized Poisson equation.  Now we investigate the influence of residual errors.  Using the same initial conditions from \S\ref{sec:polytropes}, we evolve the polytropes with overlying atmospheres using (a) the ``gravity flux" scheme with $\mathbf{\tilde{T}_g}$, (b) the ``gravity flux" scheme with $\mathbf{T_g}$, (c) the source-term-based scheme, (d) the source-term-based scheme with the addition of the corrective acceleration $\mathbf{g_\mathrm{corr}}$ (see \S\ref{sec:conservation}).   However, we now introduce a residual error to the gravitational potential by adding white noise with amplitude $A=10^{-4} G M /R$. 

\begin{figure}
    \centering
    \includegraphics[width=\textwidth]{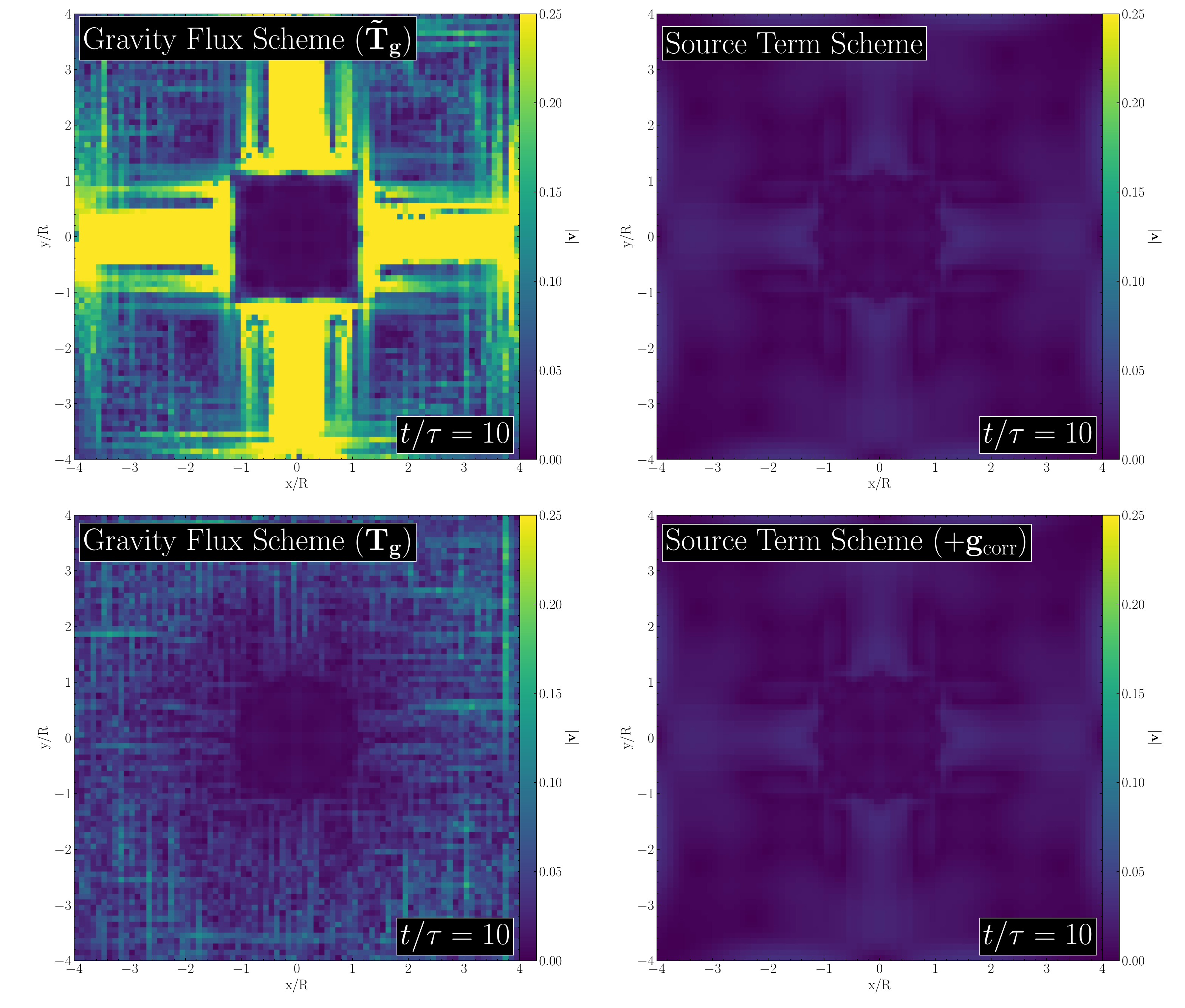}
    \caption{Equatorial velocity magnitude slices ($z=0$) of a 3-D polytrope with an overlying atmosphere after evolving the system for ten dynamical times ($t/\tau =10$) using (left) momentum ``gravity fluxes" with the gravitational stress tensor $\mathbf{\tilde{T}_g}$ (left-top) and $\mathbf{T_g}$ (left-bottom), and (right) the fully conservative source term scheme both with the corrective acceleration $\mathbf{g_\mathrm{corr}}$ (right-top) and without (right-bottom). All schemes simulate a residual error in the solution to the gravitational potential by adding white noise with amplitude $A = 10^{-4} G M/R$.}
    \label{fig:residual}
\end{figure}

Despite the introduction of the residual error, the ``gravity flux" scheme conserves total momentum to round-off error when using both $\mathbf{T_g}$ and $\mathbf{\tilde{T}_g}$.  The source term scheme does not conserve total momentum to round-off error unless the corrective acceleration $\mathbf{g_\mathrm{corr}}$ is applied.  In Figure \ref{fig:residual} we show the magnitude of the velocity in the $z=0$ plane after evolving the equilibrium for ten dynamical times ($t/\tau=10$) using the four schemes.  We find that the ``gravity flux" scheme in conjunction with $\mathbf{\tilde{T}_g}$ has large velocities corresponding to the inflows and outflows observed in \S\ref{sec:suppression}.  Although the ``gravity flux" scheme in conjunction with $\mathbf{T_g}$ gives nearly identical evolution to the source term scheme when there is no residual error, we now see differences between the two schemes in Figure \ref{fig:residual}.  The ``gravity flux" scheme with $\mathbf{T_g}$ has non-negligible velocities in low density regions that are not present in the source-term based scheme.  What are the sources of these spurious accelerations? The introduction of the residual error violates the relationship between the density $\rho_{i,j,k}$ and gravitational potential $\phi_{i,j,k}$ in the discretized Poisson equation (Equation \ref{eq:poisson_discretized2}). Hence, the force density $-(\nabla \cdot \mathbf{T_g})$ gives an implied local density that may be substantially different than $\rho_{i,j,k}$.  In contrast, the source term scheme is explicitly dependent on $\rho_{i,j,k}$ (see Equations \ref{eq:proper_momentum_discretized_x}-\ref{eq:proper_momentum_discretized_z}).  Even after the introduction of the corrective acceleration $\mathbf{g_\mathrm{corr}}$ in the source-term-based scheme, we see no evidence for the spurious behavior observed in the ``gravity flux" schemes.

\subsubsection{Energy Conservation} \label{sec:polytrope_heating}
Next, we illustrate the benefits of energy conservation when evolving self-gravitating equilibria for many dynamical times.  For this study, we compare the evolution of the 3-D polytrope (with an overlying atmosphere) using (a) the fully conservative source term scheme in this work and (b) a non-conservative scheme, where we apply energy source terms
\begin{equation} \label{eq:noncons_energy_src}
S^{(i)}_{E,i,j,k} = \left( \rho_{i,j,k}^{(i-1)} \mathbf{v}_{i,j,k}^{(i-1)} \cdot \mathbf{g}^{(i-1)} \right)_{i,j,k}.
\end{equation}
For the non-conservative energy source term in Equation (\ref{eq:noncons_energy_src}), the mass flux is estimated by multiplying the cell-centered density $\rho_{i,j,k}$ and cell-centered velocity $\mathbf{v}_{i,j,k}$.  The gravity is evaluated as in the fully conservative source term treatment, however, we perform no time averaging as in Equation (\ref{eq:proper_energy_discretized}).  We do not introduce a residual error as in \S\ref{sec:residual}.  

\begin{figure}
    \centering
    \includegraphics[width=\textwidth]{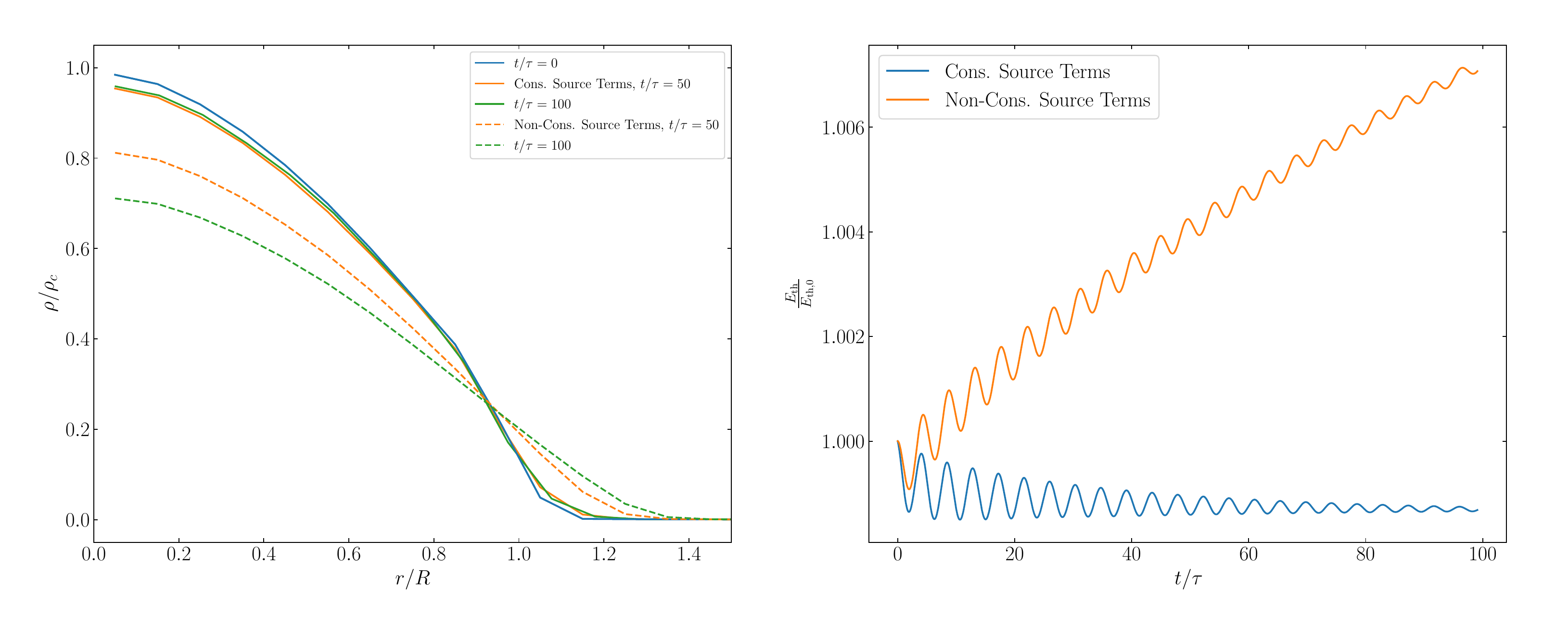}
    \caption{(Left) Spherically averaged radial density profiles for 3-D polytropic equilibria at $t/\tau = $ 0, 50, and 100 using a non-conservative scheme described in the main text and our implementation of the fully conservative source-term-based scheme. (Right) Thermal energy as a function of time for the 3-D equilibria for both the non-conservative and conservative schemes.}
    \label{fig:polytrope_energy}
\end{figure}

Figure \ref{fig:polytrope_energy} presents (left) the spherically averaged density profiles of the 3-D polytrope at $t/\tau=$0, 50, and 100 for both schemes and (right) the thermal energy as a function of time. As also observed in \cite{mikami+2008}, we find that the non-conservative source term (Equation \ref{eq:noncons_energy_src}) yields spurious heating of the polytrope, with the thermal energy growing steadily throughout the course of the run.   Expansion of the polytrope ensues.  At $t/\tau=100$, the central density has dropped by $\sim$30\%.   In the conservative scheme, the thermal energy energy shows a damping oscillation and settles into an equilibrium beyond $t/\tau \sim$20. The final state denotes the steady state solution of the numerical equilibrium (with total mass $M_\mathrm{tot} \simeq 1.0073$), which differs slightly from the analytic solution ($M_\mathrm{tot,analytic} = M + M_a = 1.01$). The small decrease observed in the thermal energy evolution is due to the slight expansion of the polytrope when reaching numerical equilibrium (partially attributed to periodic boundary conditions on the gravitational potential).

\section{Discussion} \label{sec:discussion}
Self-gravitating hydrodynamics are subject to the constraint $\nabla \times \mathbf{g} = 0$ (Equation \ref{eq:curl}).  This constraint is the analogue to the $\nabla \cdot \mathbf{B} = 0$ constraint in magnetohydrodynamics.  In a constrained transport algorithm \citep[e.g.,][]{stone+2008}, it is required that $\nabla \cdot \mathbf{B}$ vanishes, while $\nabla \times \mathbf{B}$ survives (i.e., $\nabla \times \mathbf{B} = \left( 4 \pi/c \right) \mathbf{J}$, where $c$ is the speed of light and $\mathbf{J}$ is the current density).  In a self-gravitating hydrodynamics algorithm, the situation is reversed: $\nabla \times \mathbf{g}$ vanishes, while $\nabla \cdot \mathbf{g}$ survives (i.e., $\nabla \cdot \mathbf{g} = -4 \pi G \rho$).  The implementations of the momentum ``gravity flux" in \texttt{Athena} \citep{stone+2008} and \texttt{Athena++ v19.0} \citep{stone+2020} (i.e., the discretization of the gravitational stress tensor $\mathbf{\tilde{T}_g}$) violate the $\nabla \times \mathbf{g} = 0$ constraint.  $\mathbf{\tilde{T}_g}$ approximates tangential components of the gravity as the average of the four neighboring normal components, e.g., $\mathrm{g}_{y,i+1/2,j,k} = \left(\mathrm{g}_{y,i+1,j+1/2,k} + \mathrm{g}_{y,i,j+1/2,k} + \mathrm{g}_{y,i+1,j-1/2,k} + \mathrm{g}_{y,i,j-1/2,k} \right)/4$.  Such averages are used for the tangential gravity in all components of $\mathbf{\tilde{T}_g}$.  We have identified a new discretization of the gravitational stress tensor $\mathbf{T_g}$ that gives a curl-free gravity and only differs from $\mathbf{\tilde{T}_g}$ in diagonal components.  For these diagonal components, the new discretization evaluates not the square of the average tangential gravity, but rather, the average of \textit{the geometric  mean of the tangential gravity squared} (see Equation \ref{eq:Tg_xx}).  These averages can be negative (near cell boundaries, where the gravity may change sign), in contrast to the square of the average tangential gravity.

Both the source term scheme and ``gravity flux" scheme with $\mathbf{\tilde{T}_g}$ will produce a gravity that contains some error due to truncation.  However, as seen in \S\ref{sec:curl_g_zero} and \S\ref{sec:polytropes}, the ``gravity flux" scheme can produce circulation around massive bodies (Figure \ref{fig:curl_g}).  Because $\nabla \times \mathbf{g}$ does not vanish, we see that an additional term enters the the divergence of the gravitational stress tensor (see Equation \ref{eq:tg_proof}).  For the test problem of maintaining a polytropric equilibrium with an overlying atmosphere, this circulation ultimately led to inflows along grid axes and outflows along grid diagonals, hence, enhancing the anisotropy about the grid origin (\S\ref{sec:suppression}).  We first observed such anomalous behavior when developing numerical models of Moon-forming giant impacts \citep{mullen+2020}.  We found that the anomalous accelerations were of such large magnitude that they destroyed (1) equilibrium initial conditions of planetary bodies with low-mass overlying atmospheres and (2) post-impact centrifugally supported debris disks (whose masses are small compared to the central body).  These anomalous accelerations may affect numerical simulations of any problem with large density/mass contrasts \citep[e.g.,][]{shi+2014,shi+2016,booth2019}.

The identification of a new discretization of the stress tensor $\mathbf{T_g}$ enabled us to demonstrate that source terms, when evaluated properly, can have a corresponding flux (Appendix \ref{sec:divergence_of_the_grav_stress_tensor}).  The added algorithmic complexity to evaluate self-gravity source terms in this fully conservative manner is modest.  A comfortable implementation of the scheme may require additional memory to store the gravity at the initial stage, or for the RK2 and RK3 integrators (see Appendices \ref{sec:rk2} and \ref{sec:rk3}), to store the mass fluxes from previous stages.  For second-order temporal integrators, only two Poisson solves are required per numerical time-step (thereby not increasing the number of Poisson solves compared to \texttt{Athena++ v19.0}). The scheme does not require an elliptic solve to set $\dot{\phi}$ as in the evaluation of $\mathbf{F_g}$ in the conservative scheme of \cite{jiang+2013}.  The fully conservative source term scheme requires executing a Poisson solve after the evolution of the continuity equation but before the application of the gravitational energy release source term.  In our experience, this may be the largest hurdle in adapting numerical magnetohydrodynamics software to use the proposed scheme.  

\section{Conclusion} \label{sec:conclusion}
We have shown that self-gravity source terms, when properly evaluated, are capable of being fully conservative \citep[cf.,][]{mikami+2008,springel2010,katz+2016,hanawa2019}.  The source terms are derived by guaranteeing their equivalence to a corresponding flux (i.e., for momentum source terms, the divergence of the gravitational stress tensor, and for the energy source term, the divergence of a gravitational energy flux).  As presented in this work, the fully conservative source term scheme is formally second-order accurate in space and time and does not increase the total number of Poisson solves needed per numerical time-step compared to \texttt{Athena++ v19.0}.  The scheme can be implemented alongside a broad class of temporal integrators (e.g., VL2, RK2, RK3, etc.), Riemann solvers, or reconstruction methods (e.g., PLM, PPM, etc.).  

The three test problems presented in \S\ref{sec:implementation_in_athena++} exemplify the key advantages of the fully conservative source term scheme.  For the 1-D Spitzer sheet advection test problems, we see that the the $L_1$ errors are nearly identical when the momentum equation is integrated via the divergence of the gravitational stress tensor $\mathbf{T_g}$ and the conservative source terms. The similarities in the errors and error convergence between the two Spitzer sheet runs are a reflection of the two schemes being equivalent via Equations (\ref{eq:proper1}) and (\ref{eq:proper2}).

The Jeans linear stable wave tests demonstrate that the scheme can be used in conjunction with a variety of temporal integrators (e.g., VL2, RK2, and RK3) and reconstruction methods (e.g., PLM and PPM).  Although we use the HLLC Riemann solver for all test problems, any solver can be employed; the scheme only mandates that the energy source term uses the same mass fluxes applied in evolving the continuity equation (i.e., the energy source term must be consistent with mass conservation).  The source terms in \S\ref{sec:proper_src} only guarantee spatial second-order accuracy, however, the Jeans linear wave analysis shows that higher order temporal integrators and reconstruction methods are still capable of significantly lowering errors at a given resolution.  

The Jeans instability test demonstrates the algorithm's ability to conserve total energy. Despite significant changes in each component of the volume-integrated energy, total energy is conserved to round-off error.  Total energy conservation (to round-off error) requires that the solution to the discretized Poisson equation (Equation \ref{eq:poisson_discretized}) is of round-off error accuracy at the beginning and final stages of the numerical time-step.  As noted in \S\ref{sec:conservation}, the algorithm does not require machine-accurate evaluations of the potential at intermediate stages for total energy conservation.  This could yield interesting strategies when solving the Poisson equation with iterative methods, i.e., perhaps, the convergence threshold could be relaxed at intermediate stages and made stricter at initial and final stages.  Nevertheless, as noted in \cite{katz+2016}, departures from round-off error energy conservation due to residual errors in the gravitational potential may be negligible in comparison to contributions from other common numerical effects, e.g., density/pressure floors or temperature/velocity ceilings.

The test problems evolving 3-D polytropes with overlying atmospheres show that the fully conservative scheme is capable of maintaining equilibria for many dynamical times, a requirement relevant to many astrophysical simulations (e.g., in modeling low-mass disks/atmospheres around stars/planets).  The fully conservative source-term-based scheme is not plagued by the anomalous accelerations exhibited by the numerical discretization of the gravitational stress tensor $\mathbf{\tilde{T}_g}$ (see \S\ref{sec:curl_g_zero}), nor spurious heating exhibited by non-conservative approaches (e.g., Equation \ref{eq:noncons_energy_src}).  In Figure \ref{fig:polytrope} (left), the momentum ``gravity flux", as implemented in \texttt{Athena++ v19.0}, destroys the $M_a/M = 10^{-2}$ equilibrium atmosphere of the polytrope within a few dynamical times, leaving behind an $m=4$ structure in the aftermath.  The problem is exacerbated for even lower mass atmospheres (see \S\ref{sec:curl_g_zero}).  The non-conservative source term in \S\ref{sec:polytrope_heating} destroyed the equilibrium polytrope in tens of dynamical times, whereas the fully conservative scheme maintained the equilibrium quite well for more than $\sim$100 dynamical times.  

We have not yet implemented the scheme alongside an adaptive mesh refinement (AMR) framework in \texttt{Athena++}, however, \cite{hanawa2019} shows that the scheme is straightforwardly extendable.  This issue will be revisited when multigrid AMR is available in \texttt{Athena++} (Tomida et al., in preparation).  

\acknowledgments

P. D. Mullen and C. F. Gammie are supported by the National Aeronautics and Space Administration under Grant Award 80NSSC19K0515 issued through the Emerging Worlds Program. T. Hanawa is supported by JSPS KAKENHI Grant Number JP19K03906.  We gratefully acknowledge the \texttt{yt} project \citep{turk+2011}, which made many of the visualizations in this work possible.  We thank Jonah Miller, Ben Ryan, George Wong and the \texttt{Athena++} collaboration (particularly Matt Coleman, Chang-Goo Kim, Sanghyuk Moon, and Jim Stone) for their help and comments.   We thank the anonymous referee for a careful review that improved the paper.  We gratefully acknowledge supercomputer time on NASA’s Pleiades (allocation HEC-SMD-18-1885), TACC’s stampede2 at the University of Texas at Austin (allocation TG-AST170024), and NCSA’s Blue Waters at the University of Illinois at Urbana-Champaign (allocation ILL\_bawj).


\vspace{5mm}
\facilities{Pleiades, XSEDE (stampede2), Blue Waters}


\software{Athena++ \citep{stone+2020}, yt \citep{turk+2011}
          }



\newpage 

\appendix
\section{Divergence of the Gravitational Stress Tensor} 
\label{sec:divergence_of_the_grav_stress_tensor}
\begin{figure}
    \centering
    \includegraphics[width=\textwidth]{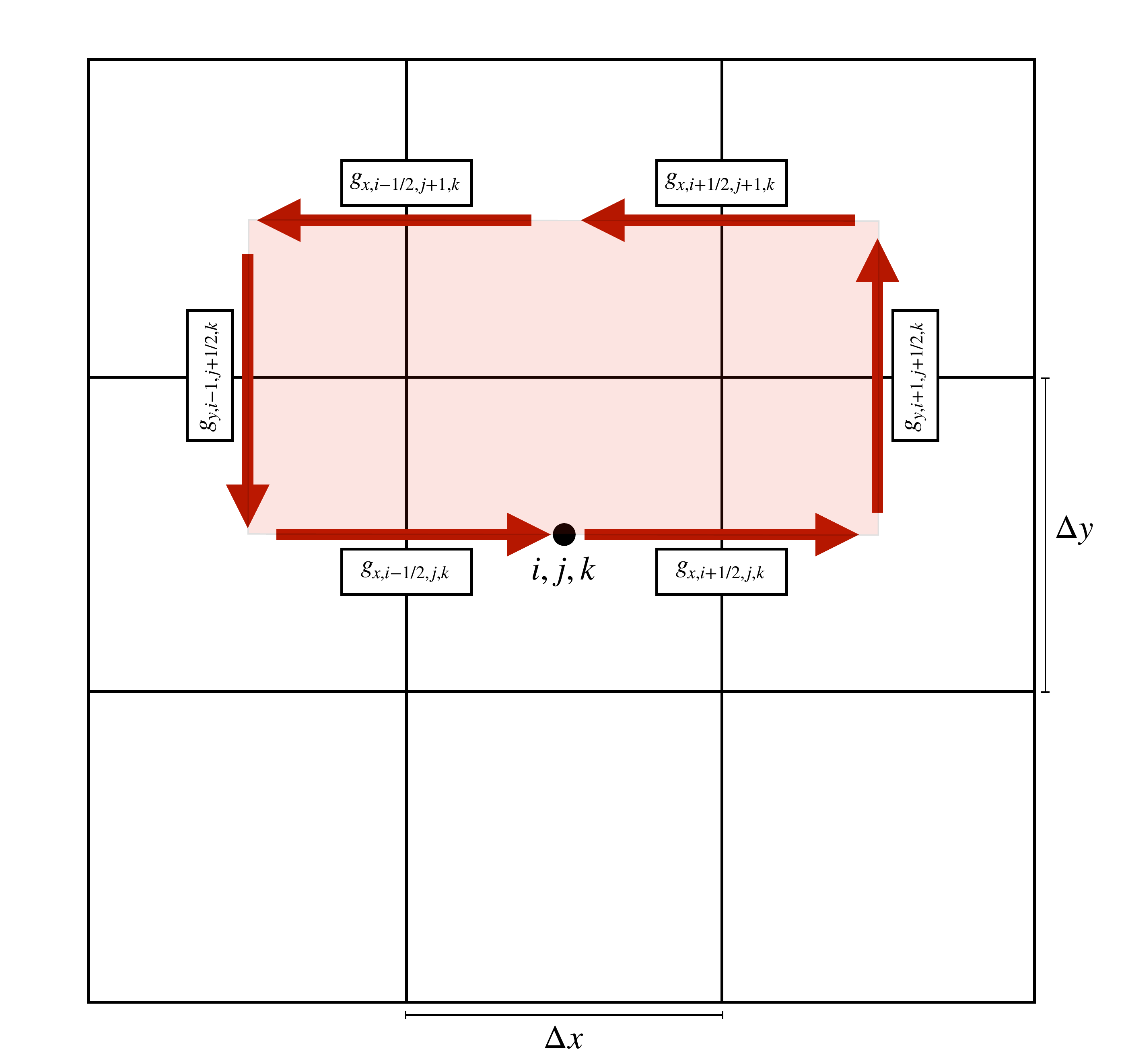}
    \caption{Schematic of a slice through a 3-D Cartesian mesh (i.e, fixed $k$) showing a subset of the face-centered, normal components of $\mathbf{g}$. The closed loop used in deriving Equation (\ref{eq:closed_loop}) is shown in red.}
    \label{fig:loop}
\end{figure}

We show that Equations (\ref{eq:proper_momentum_discretized_x}), (\ref{eq:proper_momentum_discretized_y}), and (\ref{eq:proper_momentum_discretized_z}) follow from the computation of the divergence of the gravitational stress tensor $\mathbf{T_g}$.  Equation (\ref{eq:proper1}) gives 
\begin{eqnarray}
 -\left( \nabla \cdot \mathbf{T_g} \right)_{x,i,j,k} = - \frac{T _{xx,i+1/2,j,k} - T _{xx,i-1/2,j,k}}{\Delta x} & - & \frac{T _{yx,i,j+1/2,k} -T _{yx,i,j-1/2,k}}{\Delta y} \\
 - \frac{T _{zx,i,j,k+1/2} -T _{zx,i,j,k-1/2}}{\Delta z}
 \label{eq:divTg}
\end{eqnarray}
where
\begin{eqnarray}
\frac{T _{xx,i+1/2,j,k} - T _{xx,i-1/2,j,k}}{\Delta x}  & = &
\frac{\left( \mathrm{g} _{x,i+1/2,j,k} + \mathrm{g} _{x,i-1/2,j,k} \right)}{2} \cdot \frac{\mathrm{g} _{x,i+1/2,j,k} - \mathrm{g} _{x,i-1/2,j,k}}{4 \pi G \Delta x} 
\nonumber \\
& &  - \mathrm{g} _{y,i,j+1/2,k} \cdot \frac{\mathrm{g} _{y,i+1,j+1/2,k} - \mathrm{g} _{y,i-1,j+1/2,k}}{16 \pi G \Delta x} \nonumber \\
& & - \mathrm{g} _{y,i,j-1/2,k} \cdot \frac{\mathrm{g} _{y,i+1,j-1/2,k} - \mathrm{g} _{y,i-1,j-1/2,k}}{16 \pi G \Delta x}  \nonumber \\
& &  - \mathrm{g} _{z,i,j,k+1/2} \cdot \frac{\mathrm{g} _{z,i+1,j,k+1/2} - \mathrm{g} _{z,i-1,j,k+1/2}}{16 \pi G \Delta x} \nonumber \\
& & - \mathrm{g} _{z,i,j,k-1/2} \cdot \frac{\mathrm{g} _{z,i+1,j,k-1/2} - \mathrm{g} _{z,i-1,j,k-1/2}}{16 \pi G \Delta x}  \label{eq:prior_loop_sub} \\
& = &
\frac{\left( \mathrm{g} _{x,i+1/2,j,k} + \mathrm{g} _{x,i,j+1/2,k} \right)}{2} \cdot \frac{\mathrm{g} _{x,i+1/2,j,k} - \mathrm{g} _{x,i-1/2,j,k}}{4 \pi G \Delta x} 
\nonumber \\
  - \mathrm{g} _{y,i,j+1/2,k} & \cdot & \frac{\mathrm{g} _{x,i+1/2,j+1,k} + \mathrm{g} _{x,i-1/2,j+1,k}
- \mathrm{g} _{x,i+1/2,j,k} - \mathrm{g} _{x,i-1/2,j,k}}{16 \pi G \Delta y} \nonumber \\
 - \mathrm{g} _{y,i,j-1/2,k} & \cdot & \frac{\mathrm{g} _{x,i+1/2,j,k} + \mathrm{g}_{x,i-1/2,j,k} - \mathrm{g} _{x,i+1/2,j-1,k} - \mathrm{g} _{x,i-1/2,j-1,k}}{16 \pi G \Delta y}  \nonumber \\
- \mathrm{g} _{z,i,j,k+1/2} & \cdot & \frac{\mathrm{g} _{x,i+1/2,j,k+1} + \mathrm{g} _{x,i-1/2,j,k+1} - \mathrm{g} _{x,i+1/2,j,k}
- \mathrm{g} _{x,i-1/2,j,k}}{16 \pi  G \Delta z} \nonumber \\
 - \mathrm{g} _{z,i,j,k-1/2} & \cdot &  \frac{\mathrm{g} _{x,i+1/2,j,k} + \mathrm{g} _{x,i-1/2,j,k} - \mathrm{g} _{x,i+1/2,j,k-1}
- \mathrm{g} _{x,i-1/2,j,k-1}}{16 \pi G \Delta z}, \label{eq:dTxx} \\
\frac{T _{yx,i,j+1/2,k} -T _{yx,i,j-1/2,k}}{\Delta y} 
& = & \frac{\left( \mathrm{g} _{x,i+1/2,j,k} + \mathrm{g} _{x,i-1/2,j,k} \right)}{2} \cdot
\frac{\mathrm{g} _{y,i,j+1/2,k} - \mathrm{g} _{y,i,j-1/2,k}}{4 \pi G \Delta y} \nonumber \\
+ \mathrm{g} _{y,i,j+1/2,k} & \cdot & \frac{\mathrm{g} _{x,i+1/2,j+1,k} + \mathrm{g} _{x,i-1/2,j+1,k}
- \mathrm{g} _{x,i+1/2,j,k} - \mathrm{g} _{x,i-1/2,j,k}}{16 \pi G \Delta y} \nonumber \\
+ \mathrm{g} _{y,i,j-1/2,k} & \cdot & \frac{\mathrm{g} _{x,i+1/2,j,k} + \mathrm{g}_{x,i-1/2,j,k} - \mathrm{g} _{x,i+1/2,j-1,k} - \mathrm{g} _{x,i-1/2,j-1,k}}
{16 \pi G \Delta y} \label{eq:dTyx} \\
\frac{T _{zx,i,j,k+1/2} -T _{zx,i,j,k-1/2}}{16 \pi G \Delta z} 
& = & \frac{\left( \mathrm{g} _{x,i+1/2,j,k} + \mathrm{g} _{x,i-1/2,j,k} \right)}{2} \cdot 
\frac{\mathrm{g} _{z,i,j,k+1/2} - \mathrm{g} _{z,i,j,k-1/2}}{4 \pi G \Delta z} \nonumber \\
+ \mathrm{g} _{z,i,j,k+1/2} & \cdot & \frac{\mathrm{g} _{x,i+1/2,j,k+1} + \mathrm{g} _{x,i-1/2,j,k+1} - \mathrm{g} _{x,i+1/2,j,k}
- \mathrm{g} _{x,i-1/2,j,k}}{16 \pi G \Delta z} \nonumber \\
+ \mathrm{g} _{z,i,j,k-1/2} & \cdot &  \frac{\mathrm{g} _{x,i+1/2,j,k} + \mathrm{g} _{x,i-1/2,j,k} - \mathrm{g} _{x,i+1/2,j,k-1}
- \mathrm{g} _{x,i-1/2,j,k-1}}{16 \pi G \Delta z \label{eq:dTzx} }. 
\end{eqnarray}
In moving from Equation (\ref{eq:prior_loop_sub}) to Equation (\ref{eq:dTxx}), we have used the fact that  
\begin{eqnarray}
\oint \mathbf{g} \cdot d\mathbf{s} & = & 0 ,
    \label{eq:path_integral}
\end{eqnarray}
for any closed loop.   For illustrative purposes, Figure \ref{fig:loop} shows the closed loop deriving the relation
\begin{eqnarray}
    (
    \mathrm{g}_{y,i+1,j+1/2,k} & - &   
    \mathrm{g}_{y,i-1,j+1/2,k} ) \Delta y  = \nonumber \\
    & &
    \left(
    \mathrm{g}_{x,i+1/2,j+1,k} + 
    \mathrm{g}_{x,i-1/2,j+1,k} - 
    \mathrm{g}_{x,i+1/2,j,k} - \mathrm{g}_{x,i-1/2,j,k} \right) \Delta x
    \label{eq:closed_loop}
\end{eqnarray}
which is used in the substitution yielding the numerator in the second line of Equation (\ref{eq:dTxx}).  Similar closed loops are used in deriving the remainder of Equation (\ref{eq:dTxx}).  The smallest closed loop that can be constructed on a Cartesian mesh satisfying Equation (\ref{eq:path_integral}) connects the centers of four adjacent cells; e.g., half the loop depicted in Figure \ref{fig:loop}.
Substituting Equations (\ref{eq:dTxx}-\ref{eq:dTzx}) into Equation (\ref{eq:divTg}), we obtain 
 \begin{eqnarray}
- \left(\nabla \cdot \mathbf{T_g} \right)_{x,i,j,k}
 & = & \frac{\left( \mathrm{g} _{x,i+1/2,j,k} + \mathrm{g} _{x,i-1/2,j,k} \right)}{2} \left[-
  \frac{\mathrm{g} _{x,i+1/2,j,k} - \mathrm{g} _{x,i-1/2,j,k}}{4 \pi G \Delta x}  \right. \nonumber \\
  & - & \left.  \frac{\mathrm{g} _{y,i,j+1/2,k} - \mathrm{g} _{y,i,j-1/2,k}}{4 \pi G \Delta y} 
  - \frac{\mathrm{g} _{z,i,j,k+1/2} - \mathrm{g}_{z,i,j,k-1/2}}{4 \pi G \Delta z}  \right] \label{eq:prior_poisson_sub}
  \end{eqnarray}
Substituting the discretized Poisson equation (Equation \ref{eq:poisson_discretized2}) into Equation (\ref{eq:prior_poisson_sub}), we find
\begin{eqnarray}
 - \left(\nabla \cdot \mathbf{T_g} \right)_{x,i,j,k} = \rho _{i,j,k}  \cdot \frac{\left( \mathrm{g} _{x,i+1/2,j,k} + \mathrm{g} _{x,i-1/2,j,k} \right)}{2}. \label{eq:conclude_divTgx_proof}
\end{eqnarray}
Equations (\ref{eq:proper_momentum_discretized_y}) and (\ref{eq:proper_momentum_discretized_z})
are similarly proved via the $y$- and $z$-components of $-\left( \nabla \cdot \mathbf{T_g} \right)$.

\section{Extension to Runge-Kutta Type Integrators} \label{sec:ext_to_rk_type_integrators}
\subsection{RK2} \label{sec:rk2}
First, we consider the (temporally) second order accurate, two stage RK2 integrator, otherwise known as Heun's method \citep{gottlieb+2009}.  We denote conservative variables at the initial stage, intermediate stage, and final stage as $\mathbf{U}^{(0)}$, $\mathbf{U}^{(1)}$, and $\mathbf{U}^{(2)}$.  Heun's method gives 
\begin{eqnarray}
\mathbf{U}^{(1)} & = & \mathbf{U}^{(0)} + \Delta t \mathbf{L} \left[\mathbf{U}^{(0)} \right], \\ 
\mathbf{U}^{(2)} & = & \frac{1}{2} \mathbf{U}^{(0)} + \frac{1}{2} \left\{\mathbf{U}^{(1)} + \Delta t \mathbf{L} \left[\mathbf{U}^{(1)}  \right] \right\},
\end{eqnarray}
where $\mathbf{L} \left[\mathbf{U} \right]$ denotes the operator for (magneto)hydrodynamic time marching computed from $\mathbf{U}$.

The densities at the intermediate and final stages are expressed as,
\begin{eqnarray}
\rho ^{(1)} & = & \rho ^{(0)} - \Delta t \mathbf{\nabla} \cdot \left\{ \mathcal{F_\rho} \left[\mathbf{U^{(0)}} \right] \right\} , 
\label{eq:initial_rk2}
\\
\rho ^{(2)} & = & \frac{1}{2} \rho ^{(0)} + \frac{1}{2}
\left( \rho^{(1)} - \Delta t \mathbf{\nabla} \cdot \mathcal{F_\rho} \left[\mathbf{U}^{(1)} \right]  \right) \\
& = & \rho ^{(0)} - \Delta  t \mathbf{\nabla} \cdot \left\{ \frac{ 
\mathcal{F_\rho} \left[\mathbf{U}^{(0)} \right]  + \mathcal{F_\rho} \left[\mathbf{U}^{(1)} \right]}{2}  \right\}, \label{eq:final_rk2}
\end{eqnarray}
where $\mathcal{F_\rho} \left[\mathbf{U} \right]$ is the Riemann mass flux computed from reconstructed $\mathbf{U}$.  

The momentum source terms follow
\begin{eqnarray}
S^{(1)}_{\rho \mathbf{v},i,j,k} & = & \left( \rho^{(0)} \mathbf{g}^{(0)} \right)_{i,j,k} \\
S^{(2)}_{\rho \mathbf{v},i,j,k} & = & \left( \rho^{(1)} \mathbf{g}^{(1)} \right)_{i,j,k}
\end{eqnarray}
where $\mathbf{g}^{(\ell)}$ is the gravity associated with the density distribution $\rho^{(\ell)}$ and right hand sides are evaluated following Equations (\ref{eq:proper_momentum_discretized_x}-\ref{eq:proper_momentum_discretized_z}).

The curly-braced quantities in Equations (\ref{eq:initial_rk2}) and (\ref{eq:final_rk2}) correspond to the ``effective mass fluxes",   hence, the energy source terms are 
\begin{eqnarray}
S^{(1)}_{E,i,j,k} & = & \left( \left\{ \mathcal{F_\rho} \left[\mathbf{U}^{(0)} \right] \right\} \cdot \frac{ \mathbf{g}^{(0)} + \mathbf{g}^{(1)}}{2} \right)_{i,j,k} , \\
S^{(2)}_{E,i,j,k} & = & \left( \left\{ \frac{ 
\mathcal{F_\rho} \left[\mathbf{U}^{(0)} \right]  + \mathcal{F_\rho} \left[\mathbf{U}^{(1)} \right]}{2}  \right\} \cdot \frac{ \mathbf{g}^{(0)} + \mathbf{g}^{(2)}}{2} \right)_{i,j,k}
\end{eqnarray}
where the right hand sides are evaluated following Equation (\ref{eq:proper_energy_discretized}).
We add these source terms separately at each stage, therefore, contributions from gravitational energy release from a previous intermediate stage must be removed before the addition of the new stage's gravitational release. Also, as described in \S\ref{sec:implementation_in_athena++}, the continuity equation must be evolved prior to application of the energy source terms such that we can obtain the advanced stages gravity $\mathbf{g}^{(\ell)}$ needed to compute the average gravity.

\subsection{RK3} \label{sec:rk3}
Next, we consider the (temporally) third order accurate, three stage RK3 integrator \citep{gottlieb+2009}.  Compared to the RK2 algorithm, we now must introduce a second intermediate stage.  We denote conservative variables at the initial stage, (two) intermediate stages, and final stage as $\mathbf{U}^{(0)}$, $\mathbf{U}^{(1)}$, $\mathbf{U}^{(2)}$, and $\mathbf{U}^{(3)}$.  The RK3 method follows
\begin{eqnarray}
\mathbf{U}^{(1)} & = & \mathbf{U}^{(0)} + \Delta t \mathbf{L} [\mathbf{U} ^{(0)} ] , \\
\mathbf{U}^{(2)} & = & \frac{3}{4} \mathbf{U}  ^{0} +
\frac{1}{4} \left\{\mathbf{U}^{(1)} +  \Delta t \mathbf{L} [ \mathbf{U}^{(1)} ] \right\} , \\
\mathbf{U}^{(3)} & = & \frac{1}{3} \mathbf{U}^{(0)} + 
\frac{2}{3} \left\{ \mathbf{U}^{(2)} + \Delta t \mathbf{L} [ \mathbf{U}^{(2)} ] \right\}.
\end{eqnarray}
The densities at the intermediate stages and final stage are
expressed by
\begin{eqnarray}
\rho ^{(1)} & = & \rho ^{(0)} - \Delta t \mathbf{\nabla} \cdot \mathcal{F_\rho} \left[\mathbf{U}^{(0)} \right] , \\
\rho ^{(2)} & = & \frac{3}{4} \rho ^{(0)} + \frac{1}{4}
\left\{ \rho^{(1)} - \Delta t \mathbf{\nabla} \cdot \mathcal{F_\rho} \left[\mathbf{U}^{(1)} \right] \right\} \\
& = & \rho ^{(0)} - \frac{\Delta  t}{2} \mathbf{\nabla} \cdot \left\{ 
\frac{ \mathcal{F_\rho} \left[\mathbf{U}^{(0)} \right] + \mathcal{F_\rho} \left[\mathbf{U}^{(1)} \right]}{2} \right\} , \\
\rho ^{(3)} & = & \frac{1}{3} \rho ^{(0)} + \frac{2}{3}
\left\{ \rho^{(2)} - \Delta t \mathbf{\nabla} \cdot \mathcal{F_\rho} \left[\mathbf{U}^{(2)} \right] \right\} \\
& = & \rho ^{(0)} - \Delta  t \mathbf{\nabla} \cdot \left\{\frac{
\mathcal{F_\rho} \left[\mathbf{U}^{(0)} \right] + \mathcal{F_\rho} \left[\mathbf{U}^{(1)} \right]
+ 4 \mathcal{F_\rho} \left[\mathbf{U}^{(2)} \right]}{6} \right\}.
\end{eqnarray}
The momentum source terms are
\begin{eqnarray}
S^{(1)}_{\rho \mathbf{v},i,j,k} & = & \left( \rho^{(0)} \mathbf{g}^{(0)} \right)_{i,j,k} \\
S^{(2)}_{\rho \mathbf{v},i,j,k} & = & \left( \rho^{(1)} \mathbf{g}^{(1)} \right)_{i,j,k} \\
S^{(3)}_{\rho \mathbf{v},i,j,k} & = & \left( \rho^{(2)} \mathbf{g}^{(2)} \right)_{i,j,k}.
\end{eqnarray}
The energy source terms are
\begin{eqnarray}
S^{(1)}_{E,i,j,k} & = &
\left( \left\{ \mathcal{F_\rho} \left[\mathbf{U}^{(0)} \right] \right\} \cdot 
\frac{ \mathbf{g} ^{(0)} + \mathbf{g} ^{(1)}}{2} \right)_{i,j,k}, \\
S^{(2)}_{E,i,j,k} & = &
\left( \left\{ 
\frac{ \mathcal{F_\rho} \left[\mathbf{U}^{(0)} \right] + \mathcal{F_\rho} \left[\mathbf{U}^{(1)} \right]}{2} \right\} \cdot
\frac{\mathbf{g} ^{(0)} + \mathbf{g} ^{(2)}}{2} \right)_{i,j,k}, \\
S^{(3)}_{E,i,j,k} & = &
\left( \left\{\frac{
\mathcal{F_\rho} \left[\mathbf{U}^{(0)} \right] + \mathcal{F_\rho} \left[\mathbf{U}^{(1)} \right]
+ 4 \mathcal{F_\rho} \left[\mathbf{U}^{(2)} \right]}{6} \right\} \cdot \frac{ \mathbf{g} ^{(0)} + \mathbf{g} ^{(3)} }{2} \right)_{i,j,k}.
\end{eqnarray}
Again, we add these source terms separately at each intermediate stage, removing contributions from the previous intermediate stage's source term before adding the new stage's gravitational energy release.

\bibliographystyle{aasjournal}
\bibliography{ads}

\clearpage

\clearpage

\clearpage

\clearpage

\clearpage

\clearpage

\clearpage

\end{document}